\documentclass[12pt,a4]{article}

\newcommand{\be}{\begin{equation}}
\newcommand{\ee}{\end{equation}}
\newcommand{\bee}{\begin{eqnarray}}
\newcommand{\beee}{\begin{array}}
\newcommand{\eee}{\end{eqnarray}}
\newcommand{\eeee}{\end{array}}
\newcommand{\M}{{\cal M}}

\newcommand{\ie}{{\it i.e.,} }

\newcommand{\gs}{\sigma}
\newcommand{\go}{\omega}

\newcommand{\q}{\,,\qquad}

\newcommand{\half}{\frac{1}{2}}

\usepackage{amsthm,amsmath,latexsym,multibox,amssymb,amsfonts}
\usepackage{graphicx,lscape,fancyhdr,array,stmaryrd,euscript,wrapfig}

\begin{document}
\begin{titlepage}
$$$$
\vspace{2cm}

\begin{flushright}
 FIAN/TD/20-10\\
\end{flushright}
\vspace{1cm}

\begin{center}
{\bf \Large Unfolded Scalar Supermultiplet}
\vspace{1cm}

\textsc{D.S. Ponomarev and M.A.
Vasiliev}

\vspace{.7cm}

{ I.E.Tamm Department of Theoretical Physics, P.N.Lebedev Physical
Institute,\\Leninsky prospect 53, 119991, Moscow, Russia\\
ponomarev@lpi.ru\,,\quad vasiliev@lpi.ru }

\end{center}

\vspace{0.5cm}
\begin{abstract}
Unfolded equations of motion for ${\cal N}=1$, $D=4$ scalar
supermultiplet are presented. We show how the superspace formulation
emerges from the unfolded formulation. To  analyze  supersymmetric
unfolded equations  we extend the $\sigma_-$--cohomology technics to
the case with several operators $\sigma_-$. The role of higher
$\gs_-$--cohomology in the derivation of constraints is emphasized
and illustrated by the example of the scalar supermultiplet.

\end{abstract}

\end{titlepage}

\tableofcontents

\numberwithin{equation}{section}

\section{Introduction}
\label{intro}

 A remarkable feature of the unfolded formulation of partial differential
 equations \cite{Vasilievunf1,Vasilievunf2}
 is that for the case of universal unfolded equations \cite{SolvayWorkshop},
 which in fact includes all known examples,
their form is insensitive to a particular space-time where the fields live. Since the
full system of unfolded equations keeps the same form both in space and in superspace it
is elementary to promote any supersymmetric unfolded system from, say, Minkowski space to superspace. This
idea has been already applied to the analysis of higher-spin gauge theory in superspace
\cite{Engquist:2002gy} and  generalized space-time \cite{Vasiliev:2001zy}. Although
the form of unfolded equations remains the same, its reduction to the standard
field-theoretical formulation essentially depends on the structure of background
space-time usually described by a flat connection of the symmetry group of the model
at hand. The standard machinery that uncovers the conventional field-theoretical pattern of one
or another unfolded system,  answering the questions
what are independent dynamical fields, field equations, gauge transformations etc,
is the so-called $\gs_-$ technics \cite{Shayns0}.

This property of the unfolded equations makes them convenient for the systematic
derivation (rather than guessing)  the manifestly invariant form of $G$-invariant
equations in one or another $G$-invariant space. In particular, this approach was
used in \cite{Engquist:2002gy} for superspace reformulation of nonlinear supersymmetric
higher-spin field equations and in \cite{Vasiliev:2001zy}
for the derivation of the manifestly $Sp(8)$ invariant
field equations for $4d$ massless fields reformulated in the $Sp(8)$ invariant
ten-dimensional space-time which is (the big cell of) the Lagrangian Grassmannian
$\M_4$ with local coordinates $X^{AB}=X^{BA}$ ($A,B=0,\ldots 4$ are $4D$ Majorana spinor indices).
Although the unfolded equations have similar form both in the case of
$4d$ Minkowski space and in the the ten-dimensional space $\M_4$, the patterns of
dynamical fields and field equations look differently in the two cases:
in four dimensions, the system contains an infinite set of fields of different spins,
 that satisfy massless field equations, while in $\M_4$ the same system is
described by a single scalar hyperfield $C(X)$  that satisfies
certain second-order differential field equations; similarly, all massless fermions are
described by a single spinor hyperfield $C_A(X)$, that satisfies certain first-order differential
equations. By construction, the two systems are equivalent, describing the same degrees
of freedom as was explicitly checked in \cite{Bandos:2005mb}.

The aim of this paper is to apply the methods of unfolded dynamics
\cite{Vasilievunf1, Vasilievunf2, SolvayWorkshop} to the simplest
supersymmetric model in superspace \cite{Ogievetsky:1975nu, wb, si},
namely free scalar supermultiplet in four space-time dimensions. On
the one hand, this provides an illustration of how the unfolded
dynamics approach can be used to derive superfield formulations of
supersymmetric models. On the other hand, supersymmetric models
suggest an interesting generalization of the $\gs_-$ technics at
least in two respects.

One is that usually the operator $\gs_-$, which is the negative
grade part of the covariant derivative in the unfolded field
equations, is assumed to have definite grade $-1$. We show that in
supersymmetric models it is more convenient to consider the
situation with several $\gs_-$ operators that carry different
negative grades. The corresponding generalization of the
$\gs_-$--cohomology technics suggested in this paper is rather
straightforward via application of the spectral sequence machinery
operating with cohomologies of $\gs_-$ operators of higher grades on
the cohomologies of those with lower grades.

Another comment is on the role of higher $\gs_-$--cohomologies for the derivation of
consequences of the field equations. In particular, we show how one can
use the field equations associated with higher $\gs_-$--cohomologies to distinguish
between  fundamental  field equations and their consequences.

Although application of the presented machinery to the simplest
supersymmetric model may look a bit too heavy especially in the case
of the massive model\footnote{ In fact, the unfolded formulation of
the massless case is simple enough while its massive deformation can
be reached in two ways. One is to deform fiber space constraints
that relate traces of higher grade fields to the lower grade ones.
Leading to the simpler form of equations, the analysis of the
resulting system needs independent ${\rm H}(\sigma_-)$ computation
compared to the massless case. In the scheme followed in this paper
we keep the same tracelessness fiber space constraints, modifying
instead unfolded equations which turn out to be a bit more involved.
The benefit is that the ${\rm H}(\sigma_-)$ analysis remains the
same as in the massless case.}, this is to large extent because it
contains complete information about the system under consideration,
including the structure of on-shell representation of supersymmetry
where the infinite set of zero-form fields is valued. Once obtained,
the unfolded formulation answers many questions which may be hard to
answer by other means such as, for example, the higher-spin
extension of supersymmetry of the scalar supermultiplet
\cite{Vasiliev:2001zy}.

The proposed approach can be useful for the analysis of generic
(super)symmetric theories where the $\gs_-$--cohomology analysis
makes it possible to derive systematically analogues of the Dragon
theorem \cite{Dragon:1978nf}. It should be stressed that unfolded
formulation contains all possible $G$-invariant dual formulations of
the same theory. A particular field-theoretical form depends on the
choice of one or another $G$ invariant background geometry (\ie one
or another (super)space) and on the choice of the grading which
distinguishes between dynamical and auxiliary fields in the system.
This provides a powerful tool for the analysis and classification of
$G$-invariant  dynamical systems. In particular, this approach was
used in \cite{Shaynkman:2004vu} to classify all conformal invariant
differential equations in $d$-dimensional Minkowski space-time. Once
obtained, the unfolded formulation of a given supersymmetric system
contains various its superfield formulations, allowing systematic
investigation of  all options. On the top of that unfolded
formulation provides a powerful tool for the construction of the
action and conserved currents along the lines of
\cite{Vasiliev:2005zu}. One of the most interesting problems to be
explored within the unfolded formulation in the future is the
manifestly supersymmetric off-shell action formulation of less
trivial supersymmetric systems like $N=4$ and/or $d=10$ SYM
theories.

Since the unfolded formulation proved to be most efficient for the
description of nonlinear higher-spin theories \cite{Vasiliev:1990en,
Vasiliev:2003ev, SolvayWorkshop}, their superfield formulation
should contain the description of the scalar supermultiplet in the
form presented in this paper.

The rest of the paper is organized as follows.
In Section \ref{rec}  we review the unfolded dynamics approach.
In Section \ref{gensigma} the extensions of the standard $\gs_-$ technics
 suggested by the analysis of supersymmetric models are discussed.
 In Section \ref{background} we recall the formulation of
the flat superspace in the form of the flatness conditions for
the SUSY algebra. In  Section \ref{unfeqs} we present the final result for the
 unfolded equations of a massless scalar supermultiplet with the emphasize
in its general properties such as  supersymmetry.
The $\sigma_-$--cohomology analysis of unfolded equations of Section \ref{unfeqs}
is performed in  Section \ref{sigmawz}. It is shown in particular
that the standard $\sigma_-$--cohomology technics treats the two dynamical
equations for the chiral superfield $C(z)$
\be
\label{truedynamical}
D_{\dot\alpha}C(z)=0, \qquad D^{\alpha}D_{\alpha}C(z)=0
\ee
on the same footing with their consequence
\be
\label{falsedynamical}
(\bar\sigma^a)^{\dot\alpha\alpha}D_aD_{\alpha}C(z)=0\,.
\ee
 The origin of this peculiarity in the $\sigma_-$--cohomology language as well as
 its relation to the fact that superspace possesses
  nonzero torsion are discussed in Section \ref{quest}.
In  Section \ref{nonzerom} we derive unfolded equations for  a massive scalar
supermultiplet.
 Our notations and conventions are summarized in
Appendix A. Some technicalities  are collected in Appendix B.

\section{Unfolded dynamics}
\label{rec}
 Unfolded dynamics approach \cite{Vasilievunf1,Vasilievunf2}
 implies reformulation of equations
 of motion in the  form of generalized zero curvature equations
\be
\label{1}
R^{\Omega}(x)\stackrel{def}{=}dW^{\Omega}(x)+G^{\Omega}(W(x))=0,
\ee
where $d=dx^m \frac{\partial}{\partial x^m}$ is the exterior differential
 and
\be
\notag
G^{\Omega}(W^{\Upsilon})\stackrel{def}{=}\sum_{n=1}^{\infty}f^{\Omega}_{\Upsilon_1 \dots \Upsilon_n}W^{\Upsilon_1}\dots W^{\Upsilon_n}
\ee
is built from exterior product (which is implicit in this paper)
 of differential forms
$W^{\Upsilon}(x)$ and satisfies the compatibility condition
\be
\label{3}
G^{\Upsilon}(W)\frac{\delta G^{\Omega}(W)}{\delta W^{\Upsilon}}\equiv 0\,.
\ee
Here index $\Omega$ enumerates a set
of differential forms. Let us note that (\ref{3}) is the condition
on the function $G^{\Omega}(W)$ to be satisfied identically for all $W^{\Upsilon}$.

For field-theoretical systems with infinite number of degrees of freedom,
unfolding requires  an infinite number
of auxiliary fields subjected to an infinite set of equations most of which are
algebraic constraints that express auxiliary fields via (derivatives of) dynamical fields.

The property (\ref{3}) guarantees the generalized Bianchi identity
\be
\label{B}
dR^{\Omega}=R^{\Upsilon}\frac{\delta G^{\Omega}}{\delta W^{\Upsilon}},
\ee
which tells us that the differential equations on $W^{\Upsilon}$
\be
\notag
R^{\Omega}(W)=0
\ee
are consistent with $d^2=0$.

Universal unfolded field equations are those where $W^{\Omega}$ can be treated as
coordinates of some target superspace \cite{SolvayWorkshop}.
(Alternatively, one can say that the compatibility condition (\ref{3}) holds
independently on the number of values of the indices of differential forms,
\ie it is insensitive to the fact that $p$-forms with $p>d$ are zero
in the $d$-dimensional space.)
In this case, it is possible to
differentiate freely over $W^{\Omega}$ and the equations
 (\ref{1}) are manifestly invariant under the gauge transformation
with a degree $p^\Omega -1$ differential form gauge parameter
$\varepsilon^\Omega(x)$ associated to any degree $p^\Omega >0$ form $W^\Omega$
\begin{equation}
\label{4}
\delta W^{\Omega}=d\varepsilon^{\Omega}-
\varepsilon^{\Upsilon}\frac{\delta^LG^{\Omega}(W)}{\delta W^{\Upsilon}}\,
\end{equation}
because
\begin {equation}
\notag
\delta R^{\Omega} =-R^{\Psi}\frac{\delta^L}{\delta W^{\Psi}}\left(\varepsilon^{\Upsilon}\frac{\delta^L G^{\Omega}(W)}{\delta W^{\Upsilon}}\right)\,.
\end{equation}
The gauge transformations of 0-forms $\delta C^{\Omega_0}$ only contain the gauge
parameters $\varepsilon^{\Omega_1}$ associated to 1-forms $W^{\Omega_1}$ in (\ref{4})
\begin{equation}
\label{5}
\delta C^{\Omega_0}=-\varepsilon^{\Omega_1}\frac{\delta^LG^{\Omega_0}(W)}{\delta W^{\Omega_1}}\,.
\end{equation}

In the case of universal unfolded equations, once the condition (\ref{3})
is satisfied for some base manifold, it remains consistent for any larger
(super)space. Another important fact is that in the topologically trivial situation
all information about dynamical degrees of freedom described by an unfolded system
is encoded by  0-forms $C^{\Omega_0}(x)$ at any given point $x_0$ of space-time
(see \cite{SolvayWorkshop} and references therein). Since this set of local data remains the same in any space,
 it follows that a universal unfolded system provides an
equivalent description in a larger (super)spaces simply by adding
additional coordinates corresponding to a larger (super)space.
(In this consideration it is important that the original unfolded system constitutes
a subsystem of that extended to a larger (super)space.) In particular, unfolding provides
the systematic way for derivation of constraints in superfield formulations
of supersymmetric theories.

We use the following terminology. The fields, that neither can be expressed in terms of derivatives
of other fields, nor can be gauged away are called {\it dynamical}. {\it Auxiliary fields}
are expressed via derivatives of the dynamical fields.
(As  already mentioned, unfolding usually requires an infinite
set of auxiliary fields.) Differential
conditions on the dynamical fields imposed by the unfolded equations are called
{\it dynamical equations}. Other equations are either consequences of dynamical  equations or
{\it constraints}, which are the equations satisfied identically when auxiliary fields are expressed in terms
of dynamical ones. The standard tool for the analysis of physical content
of unfolded equations is provided by $\sigma_-$--cohomology technics
which will be discussed in the next section.

An important example of unfolded equations is
\be
\label{vac}
d\Omega_0+\Omega_0\Omega_0=0,
\ee
where $\Omega_0=\Omega_0^aT_a$ is a $1$-form valued in some Lie algebra $g$
with a basis  $T_a$. The
consistency condition (\ref{3}) translates to the Jacoby identity for $g$.
Eq. (\ref{vac}) implies that the connection $\Omega_0$ is flat
which is the standard way of the description of a $g$-invariant vacuum.
This example shows how $g$-invariant background fields appear in the
unfolded equations. In the perturbative analysis, $\Omega_0$
is assumed to be of the zeroth order because it contains the background metric.

The transformation law (\ref{4}) gives the usual gauge transformations of the connection $\Omega_0$
\be
\notag
\delta\Omega_0(x)=d\varepsilon(x)+\Omega_0(x)\varepsilon(x)-\varepsilon(x)\Omega_0(x),
\ee
where $\varepsilon(x)$ is a $0$-form valued in $g$.
Given flat connection $\Omega_0(x)$ is invariant under the transformations with the
covariantly constant parameters, that satisfy
\be
\label{gs}
d\varepsilon(x)+\Omega_0(x)\varepsilon(x)-\varepsilon(x)\Omega_0(x)=0.
\ee
This equation is formally consistent by virtue of (\ref{3}). Locally, it reconstructs $\varepsilon(x)$
in terms of its value $\varepsilon(x_0)$ at any given point $x_0$. Hence, the number of
independent solutions of (\ref{gs}) coincides with
$\dim\,g$. Solutions of the equations (\ref{gs}) describe the leftover
 global symmetry $g$ of any solution of (\ref{vac}).

Let us now linearize the unfolded equation (\ref{1}) around a vacuum flat connection
$\Omega_0$, that solves (\ref{1}). To this end we set
\be
\notag
W=\Omega_0+{\cal C},
\ee
where ${\cal C}$ are differential forms of various degrees, that are treated as small
perturbations and, hence, contribute to equations linearly. Consider
the subset of ${\cal C}$ constituted by forms ${\cal C}^i_p$ of
some definite degree $p$, enumerated by index $i$.
Usually, each field ${\cal C}^i_p$ with fixed $i$ is valued in some representation of
the Lorentz-like subalgebra $h\subset g$. In the linearized approximation, one has to consider
the part of $G^i$ bilinear in $\Omega_0$ and ${\cal C}^i_p$, that is
$G^i=\Omega_0^{a}(T_a)^i{}_j{\cal C}^j_p$. In this case the condition
(\ref{3}) implies that the matrices $(T_a)^i{}_j$ form
a representation of $g$ in a vector space $V$ where the set of ${\cal C}^i_p$ with all $i$
 is valued.
The corresponding equation (\ref{1}) is the covariant constancy condition
\be
\notag
D_{\Omega_0}{\cal C}^i_p=0
\ee
with $D_{\Omega_0}\equiv d+\Omega_0$ being the covariant derivative in the $g$-module $V$.

\section{$\sigma_-$--cohomology}
\label{gensigma}
\subsection{General setup}
\label{gensigma1}

 A useful tool of the unfolding machinery is the identification of the dynamical content
 of unfolded equations with $\sigma_-$--cohomology groups
  \cite{Shayns0} (see also \cite{Vasiliev:2001zy, SolvayWorkshop}).
  The aim is to work out the dynamical pattern of the linearized unfolded system of the form
   \be
   \label{coh001}
   {\cal R}\stackrel{def}{=}(D+\sum \sigma){\cal C}=0\,,
   \ee
   where ${\cal C}$ are differential $p$-form fields, valued in $V$,  ${\cal R}$ are
 generalized curvatures,  $D$ is a covariant derivative with respect to
 Lorenz-like subalgebra $h$ of $g$ and $\sigma$ are
 operators, that act algebraically
  in the space-time sense (that is, they do not differentiate
   space-time coordinates). Some part of the equations (\ref{coh001}) has the meaning of the
    algebraic constraints that express some {\it auxiliary fields}  in terms of
    {\it dynamical fields}. The leftover equations in (\ref{coh001})
    may contain  differential equations on the dynamical fields as well as their consequences.

   Note that the decomposition of fields into auxiliary and dynamical is not necessarily
   unique. For example, in the system
      \be
   \notag
   \frac{\partial}{\partial x}B(x)+A(x)=0, \qquad \frac{\partial}{\partial x}A(x)+B(x)=0,
   \ee
   either $A$ can be interpreted as an auxiliary field and $B$
   as a dynamical field or vice versa. The resulting systems are equivalent (dual)
   to each other.

   In the $\sigma_-$--cohomology technics this ambiguity is controlled by a
   ${\mathbb Z}$ grading ${\cal G}$
that distinguishes between dynamical and auxiliary fields in such a way that auxiliary
fields have higher ${\cal G}$-grade than the respective dynamical fields.
The grading operator ${\cal G}$  is required to be diagonalizable on the space of fields and to
have  spectrum bounded
  below. Also the ${\cal G}$-grade of the exterior differential $d$ is required to be zero.
  Assuming that $h$ is the ${\cal G}$-grade zero subalgebra of $g$, the $h$ covariant
  derivative $D$ also has ${\cal G}$-grade zero. Dual formulations of the same theory
 are usually associated with different choices of the grading ${\cal G}$.
 An example of this phenomenon in higher-spin theory is considered in \cite{Matveev:2004ac}.

   Usually ${\cal G}$ counts a number of tensor indices of tensor fields of $h$, that constitute ${\cal C}$.
The $\sigma_-$--cohomology analysis applies once $\sigma$ contains a  part of negative
${\cal G}$-grade. In the case where the negative grade part
of $\sigma$ contains several operators we identify
 $\sigma_-$ with that of the lowest grade.
    Then unfolded equations (\ref{coh001}) acquire the form
   \be
   \label{coh1}
   {\cal R}\stackrel{def}{=}(D+\sigma_-+\dots){\cal C}=0,
   \ee
where ... denotes all those  operators that do not contain space-time derivatives
and have grade ${\cal G}$ higher than $\sigma_-$. The compatibility condition (\ref{3}) for (\ref{coh1})
is
\be
\label{coh01}
(D+\sigma_-+\dots)^2=0.
\ee
Decomposing this relation into different grades gives in particular that
\be
\label{s2}
 (\sigma_-)^2=0
 \ee
  since $\sigma_-$ carries the most negative grade.

 For the unfolded equations (\ref{coh1}),
the Bianchi identities (\ref{B}) and gauge transformations (\ref{4}) are
\be
   \label{coh2}
  {\cal T}\stackrel{def}{=} (D+\sigma_-+\dots){\cal R}=0\,,
   \ee
\be
   \label{coh3}
   \delta {\cal C}=(D+\sigma_-+\dots)\varepsilon.
   \ee

The $\sigma_-$--cohomology technics works as follows. Starting from the lowest grade we analyze
Eq.~(\ref{coh1}).
Fields, that are not annihilated by $\sigma_-$, can be expressed in terms of fields of lower grade
by means of (\ref{coh1}), hence they are auxiliary. The rest of fields, if cannot
be gauged away by $\sigma_-\varepsilon$ in (\ref{coh3}), are treated as dynamical. Hence,
nontrivial dynamical
fields are classified by ${\rm H}_p(\sigma_-)$. Let us note, that the space
of nontrivial dynamical $0$-form fields associated to ${\rm H}_0(\sigma_-)$
is always not empty because it at least contains the fields of the lowest grade.

Similarly, starting from the lowest grade, we impose equations (\ref{coh1}) (it
is equivalent to say, that we set associated ${\cal R}$ to zero) and analyze Bianchi identities (\ref{coh2}).
By virtue of Bianchi identities (\ref{coh2}), the part of ${\cal R}$, that is not annihilated by
$\sigma_-$,
is zero as a consequence of equations (\ref{coh1}) for lower grades, that is
${\cal R}$ is $\sigma_-$ closed.
On the other hand,
the part of the equations ${\cal R}=0$ with ${\cal R}\in {\rm Im}(\sigma_-)$ is not dynamical
 because these just impose constraints, that express
auxiliary fields in terms of (derivatives of) the fields of lower grade. We conclude that
nontrivial differential equations contained in (\ref{coh1}) are associated with
${\rm H}_{p+1}(\sigma_-)$.

Let us note that the $\sigma_-$--cohomology analysis can be naturally extended
to the bigraded or even multigraded  cases where ${\cal G}= {\mathbb Z}\times {\mathbb Z}$
 or ${\mathbb Z}^n$. Then the $\gs_-$ complex extends to a bicomplex
or multicomplex.
   For example, the ${\mathbb Z}\times {\mathbb Z}$ bicomplex structure naturally appears
in conformal supersymmetric theories \cite{Vasiliev:2001zy} as well as
   in the analysis  of partially massless \cite{Skvortsov:2006at} and
    massive \cite{Ponomarev:2010st}  higher-spin fields. In the latter  cases it
    counts the numbers of indices in the first and second rows of Young
    diagrams, associated to the tensors fields in these systems.

\subsection{Bianchi identities and consequences of dynamical equations}
\label{lowder}

Let ${\cal R}_n$ be a part of ${\cal R}$ of grade $n$ with respect to ${\cal G}$.
Suppose that by virtue of constraints and field equations ${\cal R}_n$=0 $\forall n\leq n_0$. Then,
as already mentioned in the analysis of the field equations, it follows that
those components of the curvatures ${\cal R}_{n_0+1}$ that do not belong to ${\rm Ker}(\sigma_-)$
are also zero as a consequence of Bianchi identities. The part of Bianchi identities that
is $\sigma_-$-- exact relates the curvatures ${\cal R}_{n_0+1}$ to derivatives of
${\cal R}_n$. The  part of the Bianchi
identities, that does not involve the higher curvatures ${\cal R}_{n_0+1}$ and hence may
yield nontrivial identities for dynamical equations, therefore is in ${\rm H}_{p+2}(\sigma_-)$.

The comment we wish to make in this paper is that the Bianchi identities in
${\rm H}_{p+2}(\sigma_-)$ necessarily have the trivial form $0=0$ only when $\sigma$ does not contain
parts of subleading grades. Otherwise, the Bianchi identities
may give nontrivial consequences of the field equations
that belong to ${\rm H}_{p+1}(\sigma_-)$.\footnote{One of us (MV) acknowledges
stimulating discussion of this phenomenon in a different context with Kostya Alkalaev.} In this paper we show how
this mechanism works in the the ${\cal N}=1$, $D=4$ scalar supermultiplet model. Similarly, in models where
$\sigma_-$ does not carry a definite grade, higher cohomology groups responsible for
Bianchi identities  can lead to  nontrivial consequences of field
equations. In fact, as we hope to show in more detail elsewhere, this comment translates the
Fierz-Pauli idea \cite{FP} of elimination of auxiliary fields in massive field theories
into $\sigma_-$--cohomology language. Generally, non-trivial consequences of the field
equations are characterized by ${\rm H}_{p+2}(\sigma_-)$.

Let us explain this phenomenon in some more detail.
Let the unfolded equations be of the form
\be
\label{conseq}
{\cal R}^n=D{\cal C}^n+\sigma_-{\cal C}^{n+1}+\sigma_+ \sum_{\varepsilon>0}{\cal C}^{n-
\varepsilon}=0,
\ee
where ${\cal C}^n$ is a $p$-form field of grade $n$, $\sigma_+$ is a set of operators with
positive grades and $\varepsilon>0$. The $\sigma_-$ technics yields that the dynamical equations are
\be
\label{conseq1}
(D{\cal C}^n+\sigma_+ \sum_{\varepsilon>0}{\cal C}^{n-
\varepsilon})\big|_{{\rm H}_{p+1}(\sigma_-)}=0.
\ee
 The term $\sigma_-{\cal C}^{n+1}$ does not contribute to (\ref{conseq1}) because it is
 $\sigma_-$-exact.
Fields ${\cal C}$ are expressed in terms of lower grade dynamical
 fields by means of constraints.
 Since $\varepsilon>0$, the number of
 derivatives in ${\cal C}^n$, expressed in terms of dynamical
 fields, is greater than  in  ${\cal C}^{n-\varepsilon}$.
Since ${\cal C}^n$ contributes to (\ref{conseq1}) with an additional derivative compared to
${\cal C}^{n-\varepsilon}$, the number of derivatives in the first term on the l.h.s. of
(\ref{conseq1}) is greater than the number of derivatives in the second one.

Various consequences of (\ref{conseq1}) result from its differentiations.
The problem is to find such low-derivative consequences that contain a number of
derivatives smaller than in the general case, \ie those where the highest
derivative part is zero. To achieve this,
we should find an operator, that annihilates the first  term
in (\ref{conseq1}). Let us note, that the highest derivative  term is exactly
the l.h.s. of the dynamical equations of the unfolded system of the form
\be
\label{conseq2}
\tilde{\cal R}^n=D{\cal C}^n+\sigma_-{\cal C}^{n+1}=0\,,
\ee
which can be thought of as resulting from (\ref{conseq}) in the limit where all
mass parameters are set to zero.
Hence, the problem of finding the low-derivative consequences of (\ref{conseq1}) is
equivalent to the problem of constructing  nontrivial identities for
dynamical equations of the unfolded system (\ref{conseq2}). These are given by
Bianchi identities for (\ref{conseq2}) projected to ${\rm H}_{p+2}(\sigma_-)$
\be
\label{conseq3}
(D\tilde{\cal R})\big|_{{\rm H}_{p+2}(\sigma_-)}\equiv 0.
\ee
So,  low-derivative consequences of (\ref{conseq1}) are given by
\be
\label{low}
(D{\cal R})\big|_{{\rm H}_{p+2}(\sigma_-)}= 0.
\ee

Suppose that grades of ${\rm H}_{p}(\sigma_-)$,
${\rm H}_{p+1}(\sigma_-)$ and ${\rm H}_{p+2}(\sigma_-)$ are $n_f$, $n_e$
and $n_i$ respectively.
Taking into account that $\sigma_-$ has grade $-1$,
to express auxiliary field ${\cal C}^{n_e}$ in terms
of ${\cal C}^{n_f}$
we should use $\sigma_-$--constraints $n_e-n_f$ times. This expresses
${\cal C}^{n_e}$ in terms of $(n_e-n_f)$-th derivatives of ${\cal C}^{n_f}$, which
entails that the highest order term $(D{\cal C}^{n_e})\big|_{{\rm H}_{p+1}(\sigma_-)}$
in (\ref{conseq1}) contains $n_e-n_f+1$ derivatives. Eq. (\ref{low}) has
the form $(D\sum_{\varepsilon>0}{\cal C}^{n_i-\varepsilon})\big|_{{\rm H}_{p+2}(\sigma_-)}$.
So its  highest derivative term contains $n_i-\min (\varepsilon)-n_f+1$ derivatives.
As a result, the
number of derivatives in (\ref{low}) is
less than that in (\ref{conseq1}) if $n_e-n_i+\min (\varepsilon)>0$.

\subsection{Several $\sigma_-$ operators}
\label{relcoh}

Usually, in the $\sigma_-$ analysis of unfolded equations it is assumed,
that  the  negative grade part of $\gs$ has grade $-1$. Hence, the
standard $\gs_-$--cohomology analysis considered in \cite{Shayns0, SolvayWorkshop, Skvortsov:2008vs,
 Boulanger:2008up, Skvortsov:2009nv} only treats this
particular case.
 However, as was mentioned already in \cite{Vasiliev:2001zy}
and will be explained in more detail in Section \ref{sigmawz}, in
supersymmetric models this is not the case. Hence in this subsection
we consider the peculiarities of the situation where $\sigma$
contains parts of several negative grades. In this case, usual
$\gs_-$--cohomology analysis
naturally extends to the $\gs_-$ spectral sequence analysis.

Let $\sigma'_-$ be the next to minimal negative grade, \ie
\be
\sigma = \gs_- +\gs_-^\prime +\ldots\,,
\ee
where $\ldots$ denote operators of higher grades.
Fields, equations, gauge symmetries, Bianchi identities, etc, resulting from
the $\gs_-$--cohomology analysis can be further analyzed using the operator $\sigma'_-$.
{}From (\ref{coh01}) it follows that
\be
\label{gsgs}
\{\gs_-\,,\gs_-'\} =0,
\ee
\be
\label{coh4}
(\sigma'_-)^2+\{\sigma_-,\Sigma'_-\}=0,
\ee
where $\Sigma'_-$ is the part of the operator (\ref{coh1}) of the appropriate grade,
which can in particular contain $D$.
From (\ref{gsgs}) it follows that
 $\sigma'_-$ maps  ${\rm H}(\sigma_-)$ to ${\rm H}(\sigma_-)$. In addition,
from (\ref{coh4}) it follows that $(\sigma'_-)^2=0$ when restricted to
${\rm H}(\sigma_-)$.

As a result, for
fields and curvatures that belong to ${\rm H(\sigma_-)}$,
the analysis goes along the same lines as for $\sigma_-$,
namely, to express  ${\cal C}\in{\rm H}_p(\sigma_-)$ in terms of derivatives of lower grade
fields one can use the $\sigma'_-$-constraints that belong to ${\rm H}_{p+1}(\sigma_-)$.
Analogously,
the Bianchi identities, that remain unused for expression of curvatures in terms
of lower grade ones in $\sigma_-$--cohomology analysis
belong to ${\rm H}_{p+2}(\sigma_-)$. These can be used to find the $\sigma'_-$-constraints between
eqs. (\ref{coh1}) with ${\cal R}\in {\rm H}_{p+1}(\sigma_-)$.

Let $\tilde{\sigma}'_-$ be the restriction of $\sigma'_-$ to ${\rm H}(\sigma_-)$.
In this terms, the dynamical pattern of unfolded equations is encoded by
 ${\rm H}(\tilde\sigma'_-)$ alternatively denoted as ${\rm H}(\sigma'_-| \sigma_-)$.
 Let us stress that in the analysis of the action of $\sigma'_-$ restricted to ${\rm H}(\sigma_-)$
 one should factor out all $\sigma_-$-exact terms, \ie ${\cal C}$ such that
 $\sigma'_-{\cal C}\in {\rm Im}(\sigma_-)$ belongs to ${\rm Ker}(\tilde{\sigma}'_-)$.

Analogously, one proceeds in the case where $\gs$ contains any number of
different negative grade parts, repeating the analysis of cohomologies on cohomologies
for all algebraic operators of increasing negative grades.
The more cohomologies are computed, the more relations between seemingly independent
 dynamical equations, Bianchi identities, etc are extracted.
Eventually, the pattern of the equations is governed by the cohomology
${\rm H}^p(\sigma_-'^{\dots}{}'|\dots|\sigma''_-|\sigma'_-|\sigma_-)$.
Mathematically, this translates to the spectral sequence computation.  Note that analogous
spectral sequences are familiar in computation of BRST cohomology
(see, e.g., \cite{Barnich:2004cr}).

Application of this analysis to higher cohomologies
${\rm H}^p(\sigma_-'^{\dots}{}'|\dots|\sigma''_-|\sigma'_-|\sigma_-)$,
associated to
consequences of Bianchi identities for unfolded equations that contain
$\sigma_+$ type
algebraic operators of positive grades as discussed in the previous subsection,
may help to control nontrivial consequences of dynamical equations containing
lower-derivative terms.
In Section \ref{quest} this phenomenon will be illustrated  by the $\sigma_-$ analysis of scalar
supermultiplet.

\section{Supersymmetric vacuum}
\label{background}

To start unfolding we first
 introduce gauge fields of supergravity \cite{Freedman:1976xh}
(for review see, e.g.
\cite{vanNieuwenhuizen:2004rh})
resulting from gauging the SUSY algebra that has
nonzero commutators (for notations see Appendix A)
\be
\notag
[M_{\mu\nu},M_{\rho\sigma}]=-(\eta_{\mu\rho}M_{\nu\sigma}+\eta_{\nu\sigma}M_{\mu\rho}-
\eta_{\mu\sigma}M_{\nu\rho}-\eta_{\nu\rho}M_{\mu\sigma}),
\ee
\be
\notag
[P_{\rho},M_{\mu\nu}]=\eta_{\rho\mu}P_{\nu}-\eta_{\rho\nu}P_{\mu},
\ee
\be
\notag
 \{Q_{\alpha},\bar Q_{\dot\beta}\}=-2i(\sigma^{\mu})_{\alpha\dot\beta}P_{\mu},
\ee
\be
\label{susy}
[M_{\mu\nu},Q_{\alpha}]=\frac{i}{2}(\sigma_{\mu\nu})_{\alpha}{}^{\beta}Q_{\beta}, \quad
[M_{\mu\nu},\bar Q^{\dot\alpha}]=\frac{i}{2}(\bar\sigma_{\mu\nu})^{\dot\alpha}{}_{\dot\beta}Q^{\dot\beta},
\ee
where
\be
\notag
(\sigma_{\mu\nu})_{\alpha}{}^{\beta}=\frac{1}{2}\big((\sigma_{\mu})_{\alpha\dot\alpha}(\bar\sigma_{\nu})^{\dot\alpha\beta}-(\sigma_{\nu})_{\alpha\dot\alpha}(\bar\sigma_{\mu})^{\dot\alpha\beta}\big),
\ee
\be
\notag
(\bar\sigma_{\mu\nu})^{\dot\alpha}{}_{\dot\beta}=\frac{1}{2}\big((\bar\sigma_{\mu})^{\dot\alpha\alpha}
(\sigma_{\nu})_{\alpha\dot\beta}-(\bar\sigma_{\nu})^{\dot\alpha\alpha}
(\sigma_{\mu})_{\alpha\dot\beta}\big).
\ee
 These include vierbein 1-form
$e^a=e_m{}^adx^m$, spin connection 1-form  $\omega^{a,b}=\omega_m{}^{a,b}dx^m$ and
gravitino 1-form $\phi^{\alpha}=\phi_m{}^{\alpha}dx^m$. Carrying spinorial index,
the gravitino 1-form $\phi^{\alpha}$ is  Grassmann odd.
Taking into
account that $\phi^{\alpha}$ is a degree $1$ differential form, we have the following commutation
relations for $\phi^{\alpha}$ (recall that we use the exterior products of
differential forms discarding the wedge product symbol)
\be
\notag
[\phi^{\alpha},\bar\phi^{\dot\alpha}]=0,
\quad \{\phi^{\alpha},e^a\}=0, \quad \{\phi^{\alpha},\omega^{a,b}\}=0.
\ee

Gauge fields $e^a$, $\go^{a,b}$ and $\phi^\gamma$ combine  into a 1-form
connection $\Omega$
valued in the SUSY algebra
\be
\notag
\Omega\stackrel{def}{=}e^aP_a+\frac{1}{2}\omega^{a,b}M_{ab}+\phi^{\alpha}Q_{\alpha}+\bar\phi_{\dot\alpha}
\bar Q^{\dot\alpha}\,.
\ee
A globally supersymmetric background corresponding to
the most symmetric  solution of supergravity
is described by a flat connection $\Omega$ that satisfies
\be
\label{flat}
R\stackrel{def}{=}d\Omega+\Omega\Omega=0\,.
\ee

The decomposition of the curvature $R$  into components associated to particular  generators
 \be
\notag
 R\stackrel{def}{=}S^aP_a+\frac{1}{2}R^{a,b}M_{ab}+S^{\alpha}Q_{\alpha}+\bar S_{\dot\alpha}
\bar Q^{\dot\alpha}\,
 \ee
gives the standard results
\be
\label{flat0002}
S^a=D^Le^a+2i\phi^{\alpha}\bar\phi^{\dot\alpha}(\sigma^a)_{\alpha\dot\alpha}=de^a+\omega^{a,b}e_b+2i\phi^{\alpha}\bar\phi^{\dot\alpha}(\sigma^a)_{\alpha\dot\alpha}
=0,
\ee
\be
\label{flat0003}
R^{a,b}=d\omega^{a,b}+\omega^{a,c}\omega_{c}{}^b=0,
\ee
\be
\label{flat0001}
S^{\alpha}=D^L\phi^{\alpha}=d\phi^{\alpha}+\frac
 i4\omega^{a,b}\phi^{\beta}(\sigma_{ab})_{\beta}{}^{\alpha}=
0,
\ee
\be
\label{flat1}
\bar S_{\dot\alpha}=D^L\bar\phi_{\dot\alpha}=d\bar\phi_{\dot\alpha}+\frac i4\omega^{a,b}\bar\phi_{\dot\beta}(\bar\sigma_{ab})^{\dot\beta}{}_{\dot\alpha}=
0,
\ee
where $D^L$ is the {\it Lorentz covariant derivative}. As is well-known, (\ref{flat0002})
implies  that the  torsion
$T^a=D^Le^a$ is nonzero
\be
\label{Torsion}
T^a=-2i\phi^{\alpha}\bar\phi^{\dot\alpha}(\sigma^a)_{\alpha\dot\alpha}.
\ee

As explained in Section \ref{rec}, Eq.~(\ref{flat}) has unfolded form.
Global SUSY transformations identify with those gauge (super)transformations that
leave invariant the background gauge connections in (\ref{flat0002})-(\ref{flat1}),
satisfying the equations
\be
\notag
\delta e^a=d\varepsilon^a-\varepsilon^{a,b}e_b+\varepsilon_b\omega^{a,b}-
2i(\varepsilon^{\alpha}\bar\phi^{\dot\alpha}(\sigma^a)_{\alpha\dot\alpha}+
\bar\varepsilon^{\dot\alpha}\phi^{\alpha}(\sigma^a)_{\alpha\dot\alpha})=0,
\ee
\be
\notag
\delta \omega^{a,b}=d \varepsilon^{a,b}+\omega^{a,c}\varepsilon_c{}^b-\varepsilon^{a,c}\omega_c{}^b=0,
\ee
\be
\notag
\delta\phi^{\alpha}=d\varepsilon^{\alpha}+
\frac{i}{4}\varepsilon^{\beta}\omega^{a,b}(\sigma_{ab})_{\beta}{}^{\alpha}
-\frac{i}{4}\varepsilon^{a,b}\phi^{\beta}(\sigma_{ab})_{\beta}{}^{\alpha}=0,
\ee
\be
\label{vacgauge}
\delta\bar\phi_{\dot\alpha}=d\bar\varepsilon_{\dot\alpha}+
\frac{i}{4}\bar\varepsilon_{\dot\beta}\omega^{a,b}(\bar\sigma_{ab})^{\dot\beta}{}_{\dot\alpha}
-\frac{i}{4}\varepsilon^{a,b}\bar\phi_{\dot\beta}(\bar\sigma_{ab})^{\dot\beta}{}_{\dot\alpha}=0.
\ee
Note that the flatness condition (\ref{flat}) guarantees the compatibility of the equations
(\ref{vacgauge}) which therefore reconstruct the dependence of the symmetry parameters on
space-time coordinates in terms of their values at any point of space-time.

As explained in Section \ref{rec}, to uplift the unfolded system
(which is obviously universal in the case of vacuum equations) to superspace it suffices to
add supercoordinates, $x^m\to z^A=(x^n, \theta ^\mu)$, extending appropriately the indices of differential forms:
\be
\notag
e_{m}{}^a(x)dx^m \to E_{M}{}^a(z)dz^M,  \quad \omega_m{}^{a,b}(x)dx^m \to
\Omega_M{}^{a,b}(z)dz^M,
\ee
\be
\notag
\phi_m{}^{\alpha}(x)dx^m \to E_M{}^{\alpha}(z)dz^M,
\quad \bar\phi_{m}{}^{\dot\alpha}(x)dx^m \to E_{M}{}^{\dot\alpha}(z)dz^M.
\ee
Now the zero curvature equations
\be
\label{flat2}
D^LE^a+2iE^{\alpha}E^{\dot\alpha}(\sigma^a)_{\alpha\dot\alpha}=dE^a+\Omega^{a,b}E_b+2iE^{\alpha}
E^{\dot\alpha}(\sigma^a)_{\alpha\dot\alpha}=0,
\ee
\be
\label{flatsusy1}
d\Omega^{a,b}+\Omega^{a,c}\Omega_{c}{}^b=0,
\ee
\be
\label{flatsusy2}
D^LE^{\alpha}=dE^{\alpha}+\frac
 i4\Omega^{a,b}E^{\beta}(\sigma_{ab})_{\beta}{}^{\alpha}=0,
\ee
\be
\label{flatsusy3}
D^LE_{\dot\alpha}=d\bar E_{\dot\alpha}+\frac i4\Omega^{a,b}\bar E_{\dot\beta}(\bar\sigma_{ab})^{\dot\beta}{}_{\dot\alpha}=0
\ee
describe flat superspace.
Fields $E^a$, $E^{\alpha}$ and $E_{\dot\alpha}$ can be
combined into supervierbein $E^A=dz^ME_M{}^A$.
A particular solution of (\ref{flat2})-(\ref{flatsusy3}), which extends
 Cartesian coordinates to superspace is
\be
\notag
E^a=dx^m\delta_m{}^a+d\theta^\mu \big(i\bar\theta^{\dot\mu}(\sigma^a)_{\mu\dot\mu}\big)
+d\bar\theta_{\dot\mu}\big(i\theta^{\mu}(\sigma^a)_{\mu\dot\nu}\varepsilon^{\dot\mu\dot\nu}\big),
\ee
\be
\label{cart}
E^{\alpha}=d\theta^{\mu}\delta_{\mu}{}^{\alpha}, \quad
 E_{\dot\alpha}=d\bar\theta_{\dot\mu}\delta^{\dot\mu}{}_{\dot\alpha}, \quad \Omega^{a,b}=0.
\ee

The Lorentz covariant derivative in the superspace can be rewritten in the form
\be
\label{dectangent}
D^L=E^aD_a+E^{\alpha}D_{\alpha}+E_{\dot\alpha}\bar D^{\dot\alpha}.
\ee
In coordinates (\ref{cart}) they are
\be
\notag
D_a=\partial_a, \quad D_{\alpha}=\partial_{\alpha}-i(\sigma^{b})_{\alpha\dot\alpha}\bar{\theta}^{\dot\alpha}
\partial_{b}, \quad \bar D_{\dot\alpha}=\bar\partial_{\dot\alpha}-i{\theta}^{\alpha}(\sigma^{b})_{\alpha\dot\alpha}\partial_b.
\ee
{}From $(D^L)^2=0$ it follows  that
\be
\label{dd}
\{D_{\alpha},\bar D_{\dot\alpha}\}=2i(\sigma^a)_{\alpha\dot\alpha}D_a.
\ee

\section{Free massless scalar supermultiplet unfolded}
\label{unfeqs}

In this Section unfolded equations of motion for ${\cal N}=1$, $D=4$ free massless scalar
supermultiplet  are presented. Here we give the final result leaving details of its
derivation to Section \ref{nonzerom} where the more complicated massive case will
be considered.

We start with the unfolded equations for
 scalar supermultiplet component fields in Minkowski space. Then
we modify the equations to manifestly  supersymmetric form. Finally, we add
supercoordinates to uplift the system to superspace.

Minkowski space is described by (\ref{flat1}) with $\phi=\bar\phi=0$
\be
\label{Minkowski}
T^a=D^Le^a=0, \qquad R^{a,b}=d\omega^{a,b}+\omega^{a,c}\omega_c{}^{,b}=0.
\ee
Sometimes we will use Cartesian coordinates in the sequel
\be
\label{cartes}
e_m{}^a=\delta_m^a\q \omega_m{}^{a,b}=0\q D^L=d.
\ee

Unfolded equations for a massless scalar field $C$ in Minkowski space
 are \cite{Vasilievunf2, Shayns0}
\be
\label{s0nosusy}
R^{a(k)}\stackrel{def}{=}D^LC^{a(k)}+e_bC^{a(k)b}=0,
\ee
where $C^{a(k)} \in \mathbf{U}^k_{0}$, that is $C^{a(k)}$ are complex 0-forms valued in
symmetric traceless rank-$k$ Lorentz tensors. Introducing notation
\be
\notag
(\sigma_{\downarrow}C)^{a(k)}\stackrel{def}{=}e_bC^{a(k)b},
\ee
 (\ref{s0nosusy}) can be rewritten in the form
\be \label{s0nosusys} (D^L+\sigma_{\downarrow})C=0. \ee In these
terms, the compatibility condition (\ref{3}) requires
$(D+\sigma_{\downarrow})^2=0$ which is true by virtue of
(\ref{Minkowski}). Note that  $(\sigma_{\downarrow})^2=0$ since it contains the
antisymmetrization of two  symmetric indices when acting on $C^{a(k)}$.

Let us show following  \cite{Shayns0} that (\ref{s0nosusy}) indeed describes
a massless spin $0$ particle.
 In Cartesian coordinates (\ref{cartes}) the first two
equations of (\ref{s0nosusy}) can be rewritten in the form
\be
\label{s0nosusy1}
\partial_aC+C_a=0,
\ee
\be
\label{s0nosusy2}
\partial_bC_a+C_{ab}=0.
\ee
Substituting $C_a$ expressed in terms of $C$ from (\ref{s0nosusy1}) into (\ref{s0nosusy2}) and
taking trace we have
\be
\notag
\partial^a\partial_aC=0.
\ee
Other equations in (\ref{s0nosusy}) do not impose further conditions on $C$ just expressing
auxiliary fields $C^{a(k)}$ in terms of the dynamical field $C$.

Similarly, the equations
\be
\label{s12nosusy}
r_{\alpha}{}^{a(k)}\stackrel{def}{=}D^L\chi_{\alpha}{}^{a(k)}+e_b\chi_{\alpha}{}^{a(k)b}=0
\ee
for 0-forms  $\chi_{\alpha}{}^{a(k)} \in \mathbf{V}^k_{0}$,
that is $\chi_{\alpha}{}^{a(k)}$ are complex $0$-forms valued in symmetric traceless rank-$k$ tensor-spinors
 subjected to the   condition
 \be
 \label{sigmatrans}
 (\bar\sigma_b)^{\dot\alpha\alpha}\chi_{\alpha}{}^{a(k-1)b}=0,
 \ee
 which will be referred to as $\sigma$-transversality condition,
describe massless spin $1/2$ field in Minkowski space \cite{Vasilievunf2,Vasiliev:2004cm}.

The system of a free scalar and fermion in four dimension is
supersymmetric. However, within the above formulation supersymmetry is
not manifest. To make it manifest it suffices to extend the system (\ref{s0nosusy})
and (\ref{s12nosusy}) to the case where the connections $\phi$ and $\bar\phi$
associated with supertransformations are introduced into the equations.
Clearly,  they mix bosons and fermions.
An elementary analysis then shows that the appropriate modification is
\begin{equation}
\label{eq1}
R^{a(k)}=D^LC^{a(k)}+e_bC^{a(k)b}-\sqrt{2}\phi^{\alpha}\chi_{\alpha}{}^{a(k)}=0,
\end{equation}
\begin{equation}
\label{eq2}
r_{\alpha}{}^{a(k)}=D^L\chi_{\alpha}{}^{a(k)}+e_b\chi_{\alpha}{}^{a(k)b}-\sqrt{2}i\bar{\phi}^{\dot \alpha}(\sigma_{b})_{\alpha\dot\alpha}C^{a(k)b}=0.
\end{equation}
The formal consistency of this system relies on the flatness conditions (\ref{flat}) and
the identity
$(\sigma_b)_{\beta\dot\alpha}\chi_{\alpha}{}^{a(k)b}=
(\sigma_b)_{\alpha\dot\alpha}\chi_{\beta}{}^{a(k)b}$, which is the consequence of
the $\sigma$-transversity of $\chi$
along with the fact that spinorial indices take just two values, expressed by
the formula (\ref{rel1}).

Application of  (\ref{5}) to (\ref{eq1}), (\ref{eq2}) gives the
 gauge transformation rules of
fields $C^{a(k)}$ and $\chi_{\alpha}{}^{a(k)}$
\be
\label{eq3}
\delta C^{a(k)}=\sqrt{2}\xi^{\alpha}\chi_{\alpha}{}^{a(k)}, \quad
\delta\chi_{\alpha}{}^{a(k)}=\sqrt{2}i\bar\xi^{\dot\alpha}(\sigma_b)_{\alpha\dot\alpha}C^{a(k)b},
\ee
where $\xi^{\alpha}$ is a gauge parameter associated to $\phi^{\alpha}$.
In the Cartesian coordinates $e_m{}^a=\delta_m^a$, $D^L=d$, $\phi=\bar\phi=0$ one easily finds
from (\ref{eq1}) that $C_a=-\partial_a C$. Then,
Eq. (\ref{eq3})  yields
\be
\notag
\delta C=\sqrt{2}\xi^{\alpha}\chi_{\alpha}{}, \quad \delta\chi_{\alpha}{}=-\sqrt{2}i\bar\xi^{\dot\alpha}(\sigma_b)_{\alpha\dot\alpha}\partial^bC\,,
\ee
which is the standard supertransformation \cite{si,wb}
with the  parameter $\xi^{\alpha}$. Thus, in accordance with the general
consideration of Section \ref{rec}, that the
equations (\ref{eq1}), (\ref{eq2}) are compatible with background (\ref{flat1})
guarantees supersymmetry of the model.

The unfolded formulation of massless scalar supermultiplet in superspace
is now reached easily by adding  spinorial
supercoordinates into the same system of unfolded equations (\ref{eq1}),
(\ref{eq2})
 \begin{equation}
\label{eq4}
R^{a(k)}=D^LC^{a(k)}+E_bC^{a(k)b}-\sqrt{2}E^{\alpha}\chi_{\alpha}{}^{a(k)}=0,
\end{equation}
\begin{equation}
\label{eq5}
r_{\alpha}{}^{a(k)}=D^L\chi_{\alpha}{}^{a(k)}+E_b\chi_{\alpha}{}^{a(k)b}-\sqrt{2}iE^{\dot \alpha}(\sigma_{b})_{\alpha\dot\alpha}C^{a(k)b}=0.
\end{equation}
To show that these equations  indeed describe massless scalar supermultiplet in superspace
one should sort out independent
dynamical superfields  as well as their field equations. This is achieved via
the $\sigma_-$--cohomology analysis.

\section{Dynamical content of unfolded equations for scalar supermultiplet}
\label{sigmawz}

\subsection{Irreducible subspaces}
To perform $\sigma_-$--cohomology analysis it is useful to characterize the
pattern of the spaces, where fields and curvatures are valued, in terms of
irreducible representations of the diagonal Lorentz group, that acts both on fiber
 indices and those of differential forms.

Spaces $\mathbf{U}^k_{0}$ and $\mathbf{V}^k_{0}$, where the fields $C$ and $\chi$ are
valued,  are irreducible under Lorentz transformations for fixed $k$.
This is not the case, however, for the spaces $\mathbf{U}^k_{1}$ and
$\mathbf{V}^k_{1}$, where curvatures $R$ and $r$ are valued.
 The  differential form index can be transformed to a tangent index by the supervierbein $E_M{}^A$.
For $R_M{}^{a(k)} \in \mathbf{U}^k_{1}$ and $r_{M\alpha}{}^{a(k)} \in \mathbf{V}^k_{1}$ we have
\be
\label{transf}
R_M{}^{a(k)}=E_{MA}R^{A|a(k)}, \qquad r_{M\alpha}{}^{a(k)}=E_{MA}r_{\alpha}{}^{A|a(k)}\,.
\ee
Vertical lines in  $R^{A|a(k)}$ and $r_{\alpha}{}^{A|a(k)}$
separate the fiber indices  from those resulting from the base ones.
Decomposing the respective reducible representation of the Lorentz
group into irreducible components we obtain the following decomposition of the $1$-forms into
irreducible parts:
\be
\label{decirr1}
R=(\pi_{\downarrow}+\pi_{\uparrow}+\pi^{\times})t+(\pi_{\swarrow}+
\pi_{\searrow}+\pi_{\nwarrow}+\pi_{\nearrow})\tau,
\ee
\be
\label{decirr2}
r=(\pi_{\downarrow}+\pi_{\uparrow}+\pi_{\rightarrow}+\pi^{\times})\tau+
(\pi_{\nearrow}+\pi_{\searrow}+\pi^{\times}_{\nearrow}+\pi^{\times}_{\searrow})t,
\ee
where $t$ denotes $0$-forms without spinor indices and $\tau$ denotes $0$-forms with one spinor index,
 while the operators $\pi$ are defined up to overall factors by\footnote{Notations
  are collected in Appendix A}
\be
\notag
\pi_{\downarrow}\,: \quad  \mathbf{U}^k_{p}\rightarrow \mathbf{U}^{k-1}_{p+1},
\quad \mathbf{V}^k_{p}\rightarrow \mathbf{V}^{k-1}_{p+1},
\ee
\be
\notag
\pi_{\uparrow}\,: \quad  \mathbf{U}^k_{p}\rightarrow \mathbf{U}^{k+1}_{p+1},
\quad \mathbf{V}^k_{p}\rightarrow \mathbf{V}^{k+1}_{p+1},
\ee
\be
\notag
\pi^{\times}\,: \quad  \mathbf{U}^{k,1}_{p}\rightarrow \mathbf{U}^{k}_{p+1},
\quad \mathbf{V}^{k,1}_{p}\rightarrow \mathbf{V}^{k}_{p+1},
\ee
\be
\notag
\pi_{\searrow}\,: \quad \mathbf{\dot V}^k_{p}\rightarrow \mathbf{U}^{k}_{p+1},\quad
 \mathbf{U}^k_{p}\rightarrow \mathbf{V}^{k-1}_{p+1},
\ee
\be
\notag
\pi_{\nearrow}\,: \quad \mathbf{U}^k_{p}\rightarrow \mathbf{V}^{k}_{p+1},
\quad \mathbf{\dot V}^k_{p}\rightarrow \mathbf{U}^{k+1}_{p+1},
\ee
\be
\notag
\pi_{\rightarrow}\,: \quad \mathbf{\dot V}^k_{p}\rightarrow \mathbf{V}^{k}_{p+1},
\ee
\be
\notag
\pi_{\swarrow}\,: \quad  \mathbf{V}^k_{p}\rightarrow \mathbf{U}^{k}_{p+1},
\ee
\be
\notag
\pi_{\nwarrow}\,: \quad \mathbf{V}^k_{p}\rightarrow \mathbf{U}^{k+1}_{p+1},
\ee
\be
\notag
\pi^{\times}_{\searrow}\,: \quad  \mathbf{U}^{k,1}_{p}\rightarrow \mathbf{V}^{k-1}_{p+1},
\ee
\be
\notag
\pi^{\times}_{\nearrow}\,: \quad \mathbf{U}^{k,1}_{p}\rightarrow \mathbf{V}^{k}_{p+1}.
\ee
Explicit expressions for $\pi$ with appropriately fixed  overall factors
 are given in Appendix~B.

It is convenient to endow the spaces of (spinor-)tensors with the
{\it vertical} $\mathbb{Z}$ grading ${\cal G}$ equal to the number
$n$ of vector indices for bosons and to $n+\half$ for fermions and
{\it horizontal} $\mathbb{Z}$ grading ${\cal H}$ with three nonzero
homogeneous spaces taking  ``left", ``center" or ``right" values for
spin-tensors with dotted indices, tensors and spin-tensors with
undotted indices, respectively. In these terms, the labels of the
$\pi$-operators acquire the simple meaning. In particular, the
operators that increase or decrease ${\cal G}$-grade are endowed
with arrows pointing up or down, respectively, and similarly for the
horizontal grading. For example, an operator that maps
$\mathbf{U}^k_{p}$ to $\mathbf{V}^{k}_{p+1}$
 increases ${\cal G}$-grade by $1/2$ and maps a
 tensor to a spin-tensor with undotted index. Hence it is denoted  $\pi_{\nearrow}$.

The spaces $\mathbf{U}^{k,1}_{p}$ and $\mathbf{V}^{k,1}_{p}$, that
appear in the decomposition into irreducible parts (\ref{decirr1}),
(\ref{decirr2}) are spanned by tensors of the symmetry of the
two-row Young diagram with one cell in the second row. The
$\pi$-operators acting on $\mathbf{U}^{k,1}_{p}$ and
$\mathbf{V}^{k,1}_{p}$, that annihilate a cell in the second row of
a Young diagram, are endowed with the label $\times$.

Eqs. (\ref{eq4})-(\ref{eq5}) can be rewritten in the form:
 \begin{equation}
\label{eq6}
R=D^LC+\sigma_{\downarrow}C+\sigma_{\swarrow}\chi=0,
\end{equation}
\begin{equation}
\label{eq7}
r=D^L\chi+\sigma_{\downarrow}\chi+\sigma_{\searrow}C=0,
\end{equation}
where
\be
\notag
(\sigma_{\downarrow}C)^{a(k)}\stackrel{def}{=}E_bC^{a(k)b}, \quad  (\sigma_{\swarrow}\chi)^{a(k)}\stackrel{def}{=}-\sqrt{2}E^{\alpha}\chi_{\alpha}{}^{a(k)},
\ee
\be
\label{not1}
(\sigma_{\downarrow}\chi)_{\alpha}{}^{a(k)}\stackrel{def}{=}E_b\chi_{\alpha}{}^{a(k)b},
\quad
(\sigma_{\searrow}C)_{\alpha}{}^{a(k)}\stackrel{def}{=}-\sqrt{2}iE^{\dot \alpha}(\sigma_{b})_{\alpha\dot\alpha}C^{a(k)b}.
\ee
The operators $\sigma$ are introduced similarly to $\pi$ but the overall factors in $\sigma$
are fixed by the compatibility conditions (\ref{3}) which have the form
\be
\notag
(\sigma_{\downarrow})^2C=0, \quad (\sigma_{\downarrow})^2\chi=0,
\ee
\be
\notag
\{\sigma_{\downarrow},\sigma_{\searrow}\}C=0, \quad \{\sigma_{\downarrow},\sigma_{\swarrow}\}\chi=0,
\ee
\be
\notag
(\{D^L,\sigma_{\downarrow}\}+\sigma_{\swarrow}\sigma_{\searrow})C=0, \quad
(\{D^L,\sigma_{\downarrow}\}+\sigma_{\searrow}\sigma_{\swarrow})\chi=0,
\ee
\be
\notag
\{D^L,\sigma_{\searrow}\}C=0, \quad \{D^L,\sigma_{\swarrow}\}\chi=0.
\ee

The $\sigma$-operators possess the following $G$--grades
 \be \notag
[G,\sigma_{\downarrow}]=-\sigma_{\downarrow}, \quad
[G,\sigma_{\searrow}]=-\frac 12 \sigma_{\searrow}, \quad
[G,\sigma_{\swarrow}]=-\frac 12 \sigma_{\swarrow}. \ee In accordance
with their grades, $\sigma_{\downarrow}$ will also be denoted as
$\sigma_-^1$ and operators $\sigma_{\swarrow}$ and
$\sigma_{\searrow}$ will be combined to $\sigma_-^{1/2}$. To carry
out the analysis sketched in  Subsection {\ref{relcoh}}, we first
compute cohomology of  $\sigma_-^1$ and  then restrict
$\sigma_-^{1/2}$ to ${\rm H}(\sigma_-^1)$ to obtain
$\tilde{\sigma}_-^{1/2}$. The dynamical content of the system is
encoded by ${\rm H}(\tilde{\sigma}_-^{1/2})$.

\subsection{ ${\rm H}(\sigma_-^1)$.}

 Since only the  spaces
 $\mathbf{U}^0_{0}$ and $\mathbf{V}^0_{0}$ are annihilated by $\sigma_-^1$ and
 ${\rm Im}_0(\sigma_-^1)$ is empty,
\be
\label{fieldcoh1}
{\rm H}_0(\sigma_-^1) \cong \mathbf{U}^0_{0} \oplus  \mathbf{V}^0_{0}.
\ee

Analogously, all the 1-forms of the lowest grade
\be
\label{eqscoh1}
\mathbf{U}^0_{1}=\pi_{\downarrow}\mathbf{U}^1_{0}\oplus \pi_{\swarrow}\mathbf{V}^0_{0}
\oplus \pi_{\searrow}\mathbf{\dot V}^0_{0}
\ee
 and next to the lowest grade
\be
\label{eqscoh2}
\mathbf{V}^0_{1}=\pi_{\downarrow}\mathbf{V}^1_{0}\oplus \pi_{\rightarrow}\mathbf{\dot V}^0_{0}
\oplus \pi_{\nearrow}\mathbf{U}^0_{0}\oplus \pi_{\searrow}\mathbf{U}^1_{0}\oplus
\pi^{\times}_{\searrow}\mathbf{U}^{1,1}_{0}
\ee
 spaces are annihilated by $\sigma_-^1$. In addition, one can check, that the
 projection of $\mathbf{U}^1_{1}$
to $\pi_{\uparrow}\mathbf{U}^0_{0}$ also belongs to ${\rm Ker}(\sigma_-^1)$.
 The subspaces $\pi_{\downarrow}\mathbf{U}^1_{0}$ in (\ref{eqscoh1}) and
 $\pi_{\downarrow}\mathbf{V}^1_{0}$ in (\ref{eqscoh2}) belong to ${\rm Im}(\sigma_-^1)$. Factoring
 them out, we obtain
 \be
 \label{eqscoh3}
 {\rm H}_1(\sigma_-^1)\cong\pi_{\swarrow}\mathbf{V}^0_{0}
\oplus \pi_{\searrow}\mathbf{\dot V}^0_{0}\oplus\pi_{\rightarrow}\mathbf{\dot V}^0_{0}
\oplus \pi_{\nearrow}\mathbf{U}^0_{0}\oplus \pi_{\searrow}\mathbf{U}^1_{0}\oplus
\pi^{\times}_{\searrow}\mathbf{U}^{1,1}_{0}\oplus\pi_{\uparrow}\mathbf{U}^0_{0}.
 \ee

\subsection{ ${\rm H}(\tilde{\sigma}_-^{1/2})$.}

Since $\tilde\sigma_-^{1/2}$ is a restriction of $\sigma_-^{1/2}$ to ${\rm H}(\sigma_-^1)$,
${\cal C}\in {\rm Ker}(\tilde\sigma_-^{1/2})$ means that
$\sigma_-^{1/2}{\cal C}$ vanishes up to $\sigma_-^1$-exact terms, that is
$\sigma_-^{1/2}{\cal C}\in {\rm Im}(\sigma_-^1)$.
Since
 $\sigma_-^{1/2}\mathbf{U}^0_{0}=0$ and $\sigma_-^{1/2}\mathbf{V}^0_{0}\notin {\rm Im}(\sigma_-^1)$,
 we conclude, that
  \be
 \label{eqscoh40}
 {\rm H}_0(\tilde\sigma_-^{1/2})\cong\mathbf{U}^0_{0}.
 \ee
Hence, after all the constraints are taken into account, the only dynamical field is
 $C\in \mathbf{U}^0_{0}$:

 To analyze ${\rm H}_1(\tilde\sigma_-^{1/2})$,  first, we  drop the term
 $\pi_{\swarrow}\mathbf{V}^0_{0}$ in (\ref{eqscoh3}) which is
 $\tilde\sigma_-^{1/2}$-exact. Note that the term
 $\pi_{\searrow}\mathbf{U}^1_{0}$ is $\sigma_-^{1/2}$-exact but
 not $\sigma_-^{1/2}$-exact on ${\rm H}(\sigma_-^1)$, because $\mathbf{U}^1_{0}\notin {\rm H}(\sigma_-^1)$.

  Analogously to the case of $0$-forms, for $1$-forms we keep those of the rest terms in
  (\ref{eqscoh3}), that satisfy
 $\sigma_-^{1/2}{\cal R}\in {\rm Im}(\sigma_-^1)$. The result is
 \be
 \label{eqscoh41}
 {\rm H}_1(\tilde\sigma_-^{1/2})\cong\pi_{\searrow}\mathbf{\dot V}^0_{0}\oplus
 \pi_{\rightarrow}\mathbf{\dot V}^0_{0}\oplus\pi_{\nearrow}\mathbf{U}^0_{0}.
 \ee
Let us note, that although $\sigma_{\searrow}\pi_{\uparrow}\mathbf{U}^0_{0}$ has the form
$E^aE^{\dot\alpha}(\sigma_a)_{\alpha\dot\alpha}t$ for any $t\in \mathbf{U}^0_{0}$,
it does not belong to ${\rm Im}(\sigma_-^1)$, because, being proportional to $\sigma$-matrices,
 $E^{\dot\alpha}(\sigma_a)_{\alpha\dot\alpha}t$  does not satisfy the $\sigma$-transversality
 condition (\ref{sigmatrans}), hence
 not belonging to $\mathbf{V}^1_{1}$ or any other space where the curvatures are valued.

\subsection{Dynamical interpretation}
Let us summarize the results of the $\sigma_-$--cohomology analysis. The only dynamical superfield is
$C(z)$.
Other fields are auxiliary and  express in terms of its derivatives.
For example, the projection of $R(z)=0$ to $\pi_{\swarrow}\mathbf{V}^0_{0}$ expresses
$\chi_{\alpha}(z)$ in terms of $C(z)$:
\be
\label{exaux}
\chi_{\alpha}(z)=\frac{1}{\sqrt{2}}D_{\alpha}C(z).
\ee

Equating to zero curvature projections, that belong to (\ref{eqscoh41}), and
taking into account (\ref{exaux}) we get superfield equations
\be
\label{deq1}
R(z)\big|_{\pi_{\searrow}\mathbf{\dot V}^0_{0}}=0 \quad \Rightarrow \quad \bar D_{\dot\alpha}C(z)=0,
\ee
\be
\label{deq2}
r(z)\big|_{\pi_{\nearrow}\mathbf{U}^0_{0}}=0 \quad \Rightarrow \quad D^{\alpha}D_{\alpha}C(z)=0,
\ee
\be
\label{deq3}
r(z)\big|_{\pi_{\rightarrow}\mathbf{\dot V}^0_{0}}=0 \quad \Rightarrow \quad
 (\bar\sigma^a)^{\dot\alpha\alpha}D_aD_{\alpha}C(z)=0,
\ee
where vertical lines, that carry labels associated to some spaces,
indicate the projections to these spaces.

Obviously, (\ref{deq3}) can be obtained by the application of
$\bar D_{\dot\alpha}$ to the both sides of (\ref{deq2}), taking into account
(\ref {deq1}) and (\ref{dd}). This  means that the  $\sigma_-$--cohomology
analysis performed so far does not lead to the minimal set of independent
dynamical equations contained in the
 unfolded system (\ref{eq4}), (\ref{eq5}). As we show in the next section,
in accordance with the consideration of Subsection \ref{lowder},
the missed information on the dependence of the system of equations
(\ref{deq1})-(\ref{deq3})
follows from the analysis of higher $\gs_-$--cohomology.

\section{Extra equations from higher $\gs_-$--cohomology}
\label{quest}

Let us show that Bianchi identities relate Eq.(\ref{deq3}) to (\ref{deq1}), (\ref{deq2})
 and explain why the analysis of Subsection \ref{gensigma1} and extended analysis of
 Subsection \ref{relcoh} do not catch it. Namely, along the lines of Subsection \ref{lowder}
 it will be shown that, resulting from Bianchi identities, Eq.~(\ref{deq3})
 belongs to the projection of $D^Lr_{\alpha}$ to ${\rm H}_2(\tilde\sigma_-^{1/2})$.

The left hand side of the Bianchi identity
\be
\label{mr1}
D^Lr_{\alpha}+E_br_{\alpha}{}^b-\sqrt{2}iE^{\dot\alpha}(\sigma_b)_{\alpha\dot\alpha}R^b=0
\ee
can be decomposed into irreducible parts analogously to the case of $0$- and $1$-forms
of the previous section. For the second term this gives
\be
\label{mr2}
E_br_{\alpha}{}^b=(\pi_{\downarrow}r)_{\alpha}=(\pi_{\downarrow}(\pi_{\downarrow}+\pi_{\uparrow}+\pi_{\rightarrow}+\pi^{\times})\tau_2+\pi_{\downarrow}
(\pi_{\nearrow}+\pi_{\searrow}+\pi^{\times}_{\nearrow}+\pi^{\times}_{\searrow})t_2)_{\alpha}
\ee
(here we use the decomposition (\ref{decirr2})), while for the third term  gives
\be
\label{mr3}
E^{\dot\alpha}(\sigma_b)_{\alpha\dot\alpha}R^b=(\pi_{\searrow}R)_{\alpha}=
(\pi_{\searrow}(\pi_{\downarrow}+\pi_{\uparrow}+\pi^{\times})t_3+\pi_{\searrow}(\pi_{\swarrow}+
\pi_{\searrow}+\pi_{\nwarrow}+\pi_{\nearrow})\tau_3)_{\alpha}
\ee
(here we use the decomposition (\ref{decirr1})).

To bring the first term
\be
\label{mr4}
D^Lr_{\alpha}=(E^bD_b+E^{\beta}D_{\beta}+E_{\dot\beta}\bar D^{\dot\beta})
((\pi_{\downarrow}+\pi_{\rightarrow})\tau_1+
(\pi_{\nearrow}+\pi_{\searrow}+\pi^{\times}_{\searrow})t_1)_{\alpha}
\ee
to the desired form, $D^a$, $D^{\alpha}$ and
$\bar D_{\dot\alpha}$ should be commuted to  the operators $\pi$
in (\ref{mr4}). Most conveniently  this can be done by commuting  $D^L$ to
$\pi$ according to (\ref{flat2}). For example, let us show how this works for
the term $(\pi_{\rightarrow}\tau_1)_{\alpha}$:
\be
\label{mr5}
D^LE^b(\sigma_b)_{\alpha\dot\alpha}\tau_1^{\dot\alpha}=-2iE^{\beta}E^{\dot\beta}
(\sigma^b)_{\beta\dot\beta}(\sigma_b)_{\alpha\dot\alpha}\tau_1^{\dot\alpha}-
E^b(\sigma_b)_{\alpha\dot\alpha}(E^cD_c+E^{\gamma}D_{\gamma}+E_{\dot\gamma}\bar D^{\dot\gamma})\tau_1^{\dot\alpha}.
\ee
The first term on the r.h.s. of (\ref{mr5})  results from the
action of $D^L$ on $E^b$, which brings factor $-2iE^{\beta}E^{\dot\beta}
(\sigma^b)_{\beta\dot\beta}$ due to nonzero torsion (\ref{Torsion}) while the
second term  results from permutation of $1$-forms $D^L$ and $E^b$ bringing
a minus sign.

The important point
is that the algebraic operator $E^{\beta}E^{\dot\beta}(\sigma^b)_{\beta\dot\beta}
(\sigma_b)_{\alpha\dot\alpha}$, that
appears in (\ref{mr5}), is not
explicitly present in (\ref{eq6}), (\ref{eq7}) as an algebraic $\sigma$-operator.
Analogously to operators $\sigma_-^1$ and $\sigma_-^{1/2}$, it can be used to express
the component $\tau_1^{\dot\alpha}$ of curvature $r_{\alpha}$ in terms of other curvatures by (\ref{mr1}).
A simple calculation  shows that contraction of undotted spinor indices in the
 part of (\ref{mr1}) proportional to $E^{\alpha}E^{\dot\alpha}$ gives
\be
\label{mr6}
4i\tau_1^{\dot\alpha}+2\bar D^{\dot\alpha}t_1-(\bar\sigma_a)^{\dot\alpha\alpha}D_{\alpha}t_1^a-
2\sqrt 2i\tau_3^{\dot\alpha}=0.
\ee
This can be used to express $\tau_1^{\dot\alpha}$, which is just the l.h.s. of (\ref{deq3}),
in terms of other curvatures. It is easy to see, that (\ref{mr6}) means that the projection
of (\ref{mr1}) to $\pi_{\searrow}\pi_{\nearrow}\mathbf{\dot V}^0_{0}$ vanishes.

To see how the resulting relation (\ref{mr6}) can be derived from the
analysis of Bianchi identities in terms of higher $\gs_-$--cohomology
we observe that
\be
\label{mr7}
\pi_{\searrow}\pi_{\nearrow}\mathbf{\dot V}^0_{0} \in {\rm H}_2(\tilde\sigma_-^{1/2}),
\ee
which means, that this part of Bianchi  identity (\ref{mr1}) has not been used yet
to express  curvatures in terms of lower grade ones
 in $\sigma_-^1$ and $\tilde\sigma_-^{1/2}$--cohomology analysis. Indeed, $\sigma_-^1\pi_{\searrow}\pi_{\nearrow}\mathbf{\dot V}^0_{0}=0$ and
$\pi_{\searrow}\pi_{\nearrow}\mathbf{\dot V}^0_{0} \notin {\rm Im}(\sigma_-^1)$, so
$\pi_{\searrow}\pi_{\nearrow}\mathbf{\dot V}^0_{0}\in  {\rm H}_2(\sigma_-^1)$.
Moreover, $\sigma_-^{1/2}\pi_{\searrow}\pi_{\nearrow}\mathbf{\dot V}^0_{0}=0$
and $\pi_{\searrow}\pi_{\nearrow}\mathbf{\dot V}^0_{0}=i/\sqrt 2\sigma_-^{1/2}\pi_{\nearrow}\mathbf{\dot V}^0_{0}$, but
$\pi_{\nearrow}\mathbf{\dot V}^0_{0} \notin {\rm H}_1(\sigma_-^1)$, that proves
(\ref{mr7}).

This example provides an illustration of a  general phenomenon that higher
$\sigma_-$--co\-ho\-mo\-lo\-gy may encode nontrivial relations between field equations
resulting from the naive $\sigma_-$  analysis.

\section{Massive scalar supermultiplet}
\label{nonzerom}

In this Section we derive unfolded equations of motion for ${\cal N}=1$, $D=4$ scalar
supermultiplet of any mass and
carry out their $\sigma_-$--cohomology analysis.
We use the same set of fields as in the massless case and consider the most general form of
unfolded equations. Then, imposing compatibility conditions (\ref{3}) and fixing
some field redefinition ambiguity, we determine all the terms in equations which
will contain an arbitrary parameter of mass.

Let us introduce additional operators, that act in the following way (see Appendix B):
\be
\notag
\pi_{\swarrow}\,: \quad \mathbf{U}^k_{p}\rightarrow \mathbf{\dot V}^{k-1}_{p+1},
\ee
\be
\notag
\pi_{\nwarrow}\,: \quad \mathbf{U}^k_{p}\rightarrow \mathbf{\dot V}^{k}_{p+1},
\ee
\be
\notag
\pi_{\leftarrow}\,: \quad \mathbf{V}^k_{p}\rightarrow \mathbf{\dot V}^{k}_{p+1},
\ee
\be
\notag
\pi_{\downarrow}\,: \quad \mathbf{\dot V}^k_{p}\rightarrow \mathbf{\dot V}^{k-1}_{p+1},
\ee
\be
\notag
\pi_{\uparrow}\,: \quad \mathbf{\dot V}^k_{p}\rightarrow \mathbf{\dot V}^{k+1}_{p+1}.
\ee

The general form of equations is
\be
\label{m1}
R=DC+\sigma_{\downarrow}C+\sigma_{\uparrow}C+\sigma_{\swarrow}\chi+\sigma_{\nwarrow}\chi
+\sigma_{\searrow}\bar \chi+\sigma_{\nearrow}\bar \chi=0,
\ee
\be
\label{m2}
r=D\chi+\sigma_{\downarrow}\chi+\sigma_{\uparrow}\chi+\sigma_{\rightarrow}\bar \chi+\sigma_{\searrow}C
+\sigma_{\nearrow}C+\sigma'_{\searrow} \bar{C}
+\sigma'_{\nearrow} \bar{C}=0
\ee
and conjugated equations
\be
\label{m3}
 R^+=D \bar{C}+ \sigma_{\downarrow} \bar{C}+\sigma_{\uparrow} \bar{C}+\sigma'_{\searrow}\bar \chi+ \sigma'_{\nearrow}\bar \chi
+ \sigma'_{\swarrow}\chi+\sigma'_{\nwarrow}\chi=0,
\ee
\be
\label{m4}
\bar r=D\bar \chi+\sigma_{\downarrow}\bar \chi+\sigma_{\uparrow}\bar \chi+\sigma_{\leftarrow}\chi+ \sigma_{\swarrow} C
+\sigma_{\nwarrow}C+\sigma'_{\swarrow} \bar{C}
+\sigma'_{\nwarrow}\bar{C}=0.
\ee
Operators $\sigma$ are proportional to the corresponding operators $\pi$ with the complex
overall factors $U$, $T$, $V$ and $W$
introduced as follows (to avoid writing all indices explicitly, we denote $C^{a(k)} \in \mathbf{U}^k_{p}$
as $C(k)$, $\chi_{\alpha}{}^{a(k)} \in \mathbf{V}^k_{p}$ as $\chi(k)$ and
$\bar\chi^{\dot\alpha a(k)} \in \mathbf{V}^k_{p}$ as $\bar\chi(k)$):
\be
\notag
\sigma_{\downarrow}C(k)=U_{\downarrow}(k)\pi_{\downarrow}C(k), \quad
\sigma_{\uparrow}C(k)=U_{\uparrow}(k)\pi_{\uparrow}C(k), \quad
\sigma_{\swarrow}C(k)=U_{\swarrow}(k)\pi_{\swarrow}C(k), \quad
\ee
\be
\notag
\sigma_{\searrow}C(k)=U_{\searrow}(k)\pi_{\searrow}C(k), \quad
\sigma_{\nearrow}C(k)=U_{\nearrow}(k)\pi_{\nearrow}C(k), \quad
\sigma_{\nwarrow}C(k)=U_{\nwarrow}(k)\pi_{\nwarrow}C(k), \quad
\ee
\be
\notag
\sigma_{\downarrow}\bar{C}(k)=T_{\downarrow}(k)\pi_{\downarrow}\bar{C}(k), \quad
\sigma_{\uparrow}\bar{C}(k)=T_{\uparrow}(k)\pi_{\uparrow}\bar{C}(k), \quad
\sigma'_{\swarrow}\bar{C}(k)=T_{\swarrow}(k)\pi_{\swarrow}\bar{C}(k), \quad
\ee
\be
\notag
\sigma'_{\searrow}\bar{C}(k)=T_{\searrow}(k)\pi_{\searrow}\bar{C}(k), \quad
\sigma'_{\nearrow}\bar{C}(k)=T_{\nearrow}(k)\pi_{\nearrow}\bar{C}(k), \quad
\sigma'_{\nwarrow}\bar{C}(k)=T_{\nwarrow}(k)\pi_{\nwarrow}\bar{C}(k), \quad
\ee
\be
\notag
\sigma_{\downarrow}\chi(k)=V_{\downarrow}(k)\pi_{\downarrow}\chi(k), \quad
\sigma_{\uparrow}\chi(k)=V_{\uparrow}(k)\pi_{\uparrow}\chi(k), \quad
\sigma_{\swarrow}\chi(k)=V_{\swarrow}(k)\pi_{\swarrow}\chi(k), \quad
\ee
\be
\notag
\sigma_{\nwarrow}\chi(k)=V_{\nwarrow}(k)\pi_{\nwarrow}\chi(k), \quad
\sigma'_{\swarrow}\chi(k)=V'_{\swarrow}(k)\pi_{\swarrow}\chi(k), \quad
\sigma'_{\nwarrow}\chi(k)=V'_{\nwarrow}(k)\pi_{\nwarrow}\chi(k), \quad
\ee
\be
\notag
\sigma_{\downarrow}\bar\chi(k)=W_{\downarrow}(k)\pi_{\downarrow}\bar\chi(k), \quad
\sigma_{\uparrow}\bar\chi(k)=W_{\uparrow}(k)\pi_{\uparrow}\bar\chi(k), \quad
\sigma_{\searrow}\bar\chi(k)=W_{\searrow}(k)\pi_{\searrow}\bar\chi(k), \quad
\ee
\be
\notag
\sigma_{\nearrow}\bar\chi(k)=W_{\nearrow}(k)\pi_{\nearrow}\bar\chi(k), \quad
\sigma'_{\searrow}\bar\chi(k)=W'_{\searrow}(k)\pi_{\searrow}\bar\chi(k), \quad
\sigma'_{\nearrow}\bar\chi(k)=W'_{\nearrow}(k)\pi_{\nearrow}\bar\chi(k), \quad
\ee
\be
\label{introcoef}
\sigma_{\leftarrow}\chi(k)=V_{\leftarrow}(k)\pi_{\leftarrow}\chi(k), \quad
\sigma_{\rightarrow}\bar\chi(k)=W_{\rightarrow}(k)\pi_{\rightarrow}\bar\chi(k).
\ee

Let us note, that it is possible to add terms of the form $\pi_{\downarrow}\bar{C}$ and
$\pi_{\uparrow}\bar{C}$ to (\ref{m1}). These terms are built from vierbein $e$, and
 survive in Minkowski case with $\phi=0$. Since such terms are absent
in the unfolded equations for spin-$0$ field in Minkowski space,  we require them to
 be absent in (\ref{m1}). {In fact, such terms can be removed by a field
redefinition.}

Using that (\ref{m3}) and (\ref{m4}) are conjugated to (\ref{m1}) and (\ref{m2}), we have:
\begin{equation}
\notag
U^*_{\downarrow}(k)=T_{\downarrow}(k), \quad U^*_{\uparrow}(k)=T_{\uparrow}(k), \quad
V^*_{\swarrow}(k)=W'_{\searrow}(k),
\end{equation}
\begin{equation*}
 -V^*_{\nwarrow}(k)=W'_{\nearrow}(k), \quad W^*_{\searrow}(k)=V'_{\swarrow}(k), \quad
 -W^*_{\nearrow}(k)=V'_{\nwarrow}(k),
\end{equation*}
\begin{equation*}
V^*_{\downarrow}(k)=W_{\downarrow}(k), \quad V^*_{\uparrow}(k)=W_{\uparrow}(k), \quad
-W^*_{\rightarrow}(k)=V_{\leftarrow}(k), \quad -U^*_{\searrow}(k)=T_{\swarrow}(k),
\end{equation*}
\begin{equation}
\label{sopriazheniya}
U^*_{\nearrow}(k)=T_{\nwarrow}(k), \quad -T^*_{\searrow}(k)=U_{\searrow}(k), \quad
T^*_{\nearrow}(k)=U_{\nwarrow}(k).
\end{equation}
The compatibility conditions (\ref{3}) impose constraints on these coefficients
(\ref{introcoef}) given in Appendix B.

Eqs. (\ref{m1})-(\ref{m4}) have a freedom in  field rescaling
\be
\label{m5}
C(n)= X(n)\tilde C(n), \qquad \chi(n)=Y(n)\tilde\chi(n),
\ee
which induces the following redefinition of the coefficients $U_{\downarrow}$ and $V_{\downarrow}$
\be
\notag
\tilde U_{\downarrow}(n)=\frac{X(n)}{X(n-1)}U_{\downarrow}(n), \quad
\tilde V_{\downarrow}(n)=\frac{Y(n)}{Y(n-1)}V_{\downarrow}(n).
\ee
It is convenient to fix
\be
\label{m6}
U_{\downarrow}(n)=1, \quad V_{\downarrow}(n)=1.
\ee
The remaining  part of the rescaling ambiguity is given by (\ref{m5})
 with $X(n)=X(0)$ and $Y(n)=Y(0)$.
 Resolving the compatibility conditions in the scaling (\ref{m6}) we express all the coefficients
 \be
\notag
U_{\downarrow}(k)=V_{\downarrow}(k)=T_{\downarrow}(k)=W_{\downarrow}(k)=1,
\ee
\be
\notag
W_{\rightarrow}(k)=-B^*_{\leftarrow}\frac{1}{k+2}, \quad
V_{\leftarrow}(k)=B_{\leftarrow}\frac{1}{k+2},
\ee
\be
\notag
V_{\uparrow}(k)=W_{\uparrow}(k)=(-2B_{\leftarrow}B^*_{\leftarrow})\left(\frac{k+1}{k+2}\right)^2,
\ee
\be
\notag
U_{\uparrow}(k)=T_{\uparrow}(k)=(-2B_{\leftarrow}B^*_{\leftarrow})\frac{k+1}{k+2},
\ee
\be
\notag
U_{\searrow}(k)=A_{\searrow}, \quad U_{\nearrow}(k)= -2B^*_{\leftarrow}A_{\swarrow}, \quad T_{\swarrow}(k)=-A^*_{\searrow},
\quad T_{\nwarrow}(k)=-2B_{\leftarrow}A^*_{\swarrow},
\ee
\be
\notag
U_{\swarrow}(k)=A_{\swarrow}, \quad U_{\nwarrow}(k)= 2B_{\leftarrow}A_{\searrow}, \quad T_{\searrow}(k)=-A^*_{\swarrow},
\quad T_{\nearrow}(k)=2B^*_{\leftarrow}A^*_{\searrow},
\ee
\be
\notag
V_{\swarrow}(k)=B_{\swarrow}, \quad V_{\nwarrow}(k)= C_{\searrow}B_{\leftarrow}\frac{k+1}{k+2}, \quad
W'_{\searrow}(k)=B^*_{\swarrow}, \quad W'_{\nearrow}(k)= -C^*_{\searrow}B^*_{\leftarrow}\frac{k+1}{k+2},
\ee
\be
\label{m7}
V'_{\swarrow}(k)=C^*_{\searrow}, \quad V'_{\nwarrow}(k)= B^*_{\swarrow}B_{\leftarrow}\frac{k+1}{k+2}, \quad
W_{\searrow}(k)=C_{\searrow}, \quad W_{\nearrow}(k)= -B_{\swarrow}B^*_{\leftarrow}\frac{k+1}{k+2},
\ee
 in terms of parameters
  $B_{\leftarrow}$, $A_{\swarrow}$, $A_{\searrow}$, $B_{\swarrow}$, $C_{\searrow}$,
  subjected to the conditions
  \be
\label{param1}
C_{\searrow}A_{\swarrow}=iD_1,
\ee
\be
\label{param2}
 B_{\swarrow}A_{\searrow}=iD_2,
 \ee
 \be
\label{param3}
 A_{\searrow}C_{\searrow}=(A_{\swarrow}B_{\swarrow})^*,
\ee
\be
\label{param4}
D_1+D_2=2,
\ee
where  $D_1$ and $D_2$ are some real parameters.
From (\ref{param1})-(\ref{param4}) it follows that
\be
\label{param5}
D_1D_2\le 0.
\ee

The remaining scaling ambiguity acts on these parameters as follows
\be
\notag
\tilde B_{\swarrow}=B_{\swarrow}\frac{Y(0)}{X(0)}, \quad
\tilde C_{\searrow}=C_{\searrow}\frac{Y^*(0)}{X(0)}, \quad
\tilde A_{\searrow}=A_{\searrow}\frac{X(0)}{Y(0)}, \quad
\tilde A_{\searrow}=A_{\searrow}\frac{X(0)}{Y^*(0)}.
\ee
First, we fix
$B_{\swarrow}=-\sqrt{|D_2|}$. Then we can fix the phase of $Y(0)$
in such a way that $C_{\searrow}$ is negative and real.  Then the scaling symmetry is fixed
up to (\ref{m5}) with
\be
\label{leftsc}
X(n)=Y(n)=X(0), \quad X(0) \in \mathbb R
\ee
 and the solution
of (\ref{param1})-(\ref{param4}) is
\be
\notag
B_{\swarrow}=-\sqrt{|D_2|}, \quad A_{\searrow}=
-i\cdot{\rm sign}(D_2)\sqrt{|D_2|},
\ee
\be
\label{param}
C_{\searrow}=-\sqrt{|D_1|}, \quad  A_{\swarrow}=-i\cdot{\rm sign}(D_1)\sqrt{|D_1|}.
\ee

The curvatures (\ref{m1}), (\ref{m2}) are invariant under
 another type of symmetry. Namely, $C$  can be replaced by appropriately
normalized linear combination of $C$ and $\bar{C}$
 (the normalization is necessary to preserve (\ref{param})).
 One can show, that
\be
\notag
\tilde C(k)=\frac{1}{\sqrt{2}}\Big({\rm sign}(D_2)\sqrt{|D_2|}C(k)+{\rm sign}(D_1)\sqrt{|D_1|}\bar{C}(k)\Big)
\ee
redefines any $D_1$ and $D_2$ in such  a way that
\be
\label{param6}
D_1=0, \qquad D_2=2.
\ee
This fixes all the coefficients. The final result is given
by Eqs.~(\ref{m7}), (\ref{param}) and
(\ref{param6}).

Although the unfolded  equations for a massive supermultiplet are different from
the equations
(\ref{eq4}) and  (\ref{eq5}) for the massless case, this difference
does not affect the operators $\sigma_-$ which remain the same. Hence,
$\sigma_-$--cohomology analysis yields the same results.

The dynamical field is $C(z)$. Constraint (\ref{exaux}) and the first
dynamical equation (\ref{deq1}) remain the same. The second dynamical
equation is given by the projection of
\be
\notag
r_{\alpha}=D \chi_{\alpha}+E_b\chi^b-\frac{1}{2}B_{\leftarrow}B^*_{\leftarrow}
E^b(\sigma_b)_{\alpha\dot\alpha}\bar{\chi}^{\dot\alpha}-
\sqrt 2iE^{\dot\alpha}(\sigma_b)_{\alpha\dot\alpha}C^b+2\sqrt 2iB^*_{\leftarrow}E_{\alpha}\bar{C}=0
\ee
to $\pi_{\nearrow}\mathbf{U}^0_{0}$ and yields
\be
\label{mfinal}
0=D^{\alpha}D_{\alpha}C-8iB^*_{\leftarrow}\bar{C}.
\ee
Comparison of (\ref{mfinal}) with the standard formula gives
\be
\label{mass}
B_{\leftarrow}=\frac{i}{2}m\,.
\ee
The consistency of unfolded equations (\ref{m1})-(\ref{m4}) does not impose any constraints on
$B_{\leftarrow}$ which remains arbitrary. Formulas (\ref{deq1}) and (\ref{mfinal})
yield the standard superfield description of the scalar supermultiplet \cite{wb}:
\be
\notag
\bar D_{\dot\alpha}C(z)=0, \quad D^{\alpha}D_{\alpha}C(z)-4m\bar{C}(z)=0.
\ee

\section{Conclusion}

In this paper we derive and analyze  unfolded equations for the simplest
supersymmetric model in four space-time dimensions associated to a scalar supermultiplet.
The analysis of these equations and, more generally, other supersymmetric models,
  requires extension of the standard $\sigma_-$--cohomology technics to the case
  of unfolded equations, that contain two and more negative grade algebraic operators.
This extension is presented in this paper for a general unfolded system.
 Also we explain how to associate
the equations that result from Bianchi identities to higher $\sigma_-$--cohomology,
which is in particular necessary to extract the full information on dynamical equations
in superspace.

Since the reformulation of supersymmetric systems in the unfolded
form makes it possible to elaborate its superfield pattern in a systematic way,
it would be interesting to extend the results of this paper to the variety of other
supersymmetric models (both on-shell and off-shell) in various dimensions.
It should be noted that, in accordance with the general discussion
of Section \ref{rec}, the form of linearized unfolded equations can be systematically
derived by choosing appropriate modules of SUSY algebra to avoid a complicated
brut force analysis like that of Section \ref{nonzerom} that was possible
to use because of simplicity of the model. Also, let us mention that for
off-shell unfolded supersymmetric systems one can look for manifestly supersymmetric
actions along the lines of \cite{Vasiliev:2005zu}.

The results of this paper may have a wide area of applicability beyond
supersymmetric models. In particular, as we will explain in more detail
elsewhere, the proposed  association of higher $\sigma_-$--cohomology
with consequences of dynamical equations provides an interesting interpretation of
the old Fierz-Pauli program \cite{FP} in the massive HS models whose unfolded form was
recently given in \cite{Ponomarev:2010st}.

\section*{Acknowledgments}

The authors would like to thank Konstantin Alkalaev and Evgeny Skvortsov for
stimulating and useful discussions.
This research was supported in part by,
RFBR Grant No 08-02-00963,
 Alexander von Humboldt Foundation Grant PHYS0167
and  by Grant of UNK.

\appendix
\renewcommand{\theequation}{\Alph{section}.\arabic{equation}}
\section*{Appendix A: Notations}
\setcounter{equation}{0}
\setcounter{section}{1}

In this paper we deal with 4-dimensional space parametrized by coordinates $x^{m}$,
$(m\in \{0,1,2,3\})$. The vierbein field $e_{m}{}^a$ relates
 base indices (i.e. indices of differential forms) denoted
by Latin letters $m,n,\dots$ from the middle of the alphabet and fiber indices
(i.e. indices of tensors in local basis) denoted by Latin letters $a,b,\dots$
from the beginning of the alphabet.
In the fiber space we use the mostly minus Minkowski metric
 $\eta_{ab}=\rm {diag}(1,-1,-1,-1)$.
 In the process of unfolding we introduce auxiliary fields, which
 are traceless symmetric tensors on their fiber indices. Sometimes
we do not write  symmetrized indices explicitly, just
 writing one of them and indicating the number of indices in brackets
 (e.g., $a(k)$ instead of $(a_1\ldots a_k)$).

The fiber spinor indices are from the beginning of Greek alphabet
$\alpha, \dot\alpha, \beta, \dot\beta \dots$ $(\alpha \in \{0,1\})$, while
base spinor indices $\mu, \dot\mu, \nu, \dot\nu,
\dots$ needed for differential forms in superspace are  from the middle
of Greek alphabet. Fiber spinor
indices are raised/lowered by antisymmetric forms
\begin{equation}
\notag
\varepsilon^{\alpha\beta}=\varepsilon^{\dot\alpha\dot\beta}=
\begin{array}{||cc||}
0 & 1\\
-1 & 0
\end{array}\,, \quad
\varepsilon_{\alpha\beta}=\varepsilon_{\dot\alpha\dot\beta}=
\begin{array}{||cc||}
0 & -1\\
1 & 0
\end{array}\,,
\end{equation}
\be
\notag
\xi_{\alpha}=\varepsilon_{\alpha\beta}\xi^{\beta}, \quad \xi^{\alpha}=\varepsilon^{\alpha\beta}\xi_{\beta},
\quad \bar\chi_{\dot\alpha}=\varepsilon_{\dot\alpha\dot\beta}\bar\chi^{\dot\beta}, \quad \bar\chi^{\dot\alpha}=\varepsilon^{\dot\alpha\dot\beta}\bar\chi_{\beta}.
\ee

We use the following convention for $\sigma$-matrices
\begin{equation}
\notag
(\sigma^0)_{\alpha\dot\alpha}=
\begin{array}{||cc||}
1 & 0\\
0 & 1
\end{array}\,, \quad
(\sigma^1)_{\alpha\dot\alpha}=
\begin{array}{||cc||}
0 & 1\\
1 & 0
\end{array}\,, \quad
(\sigma^2)_{\alpha\dot\alpha}=
\begin{array}{||cc||}
0 & -i\\
i & 0
\end{array}\,, \quad
(\sigma^3)_{\alpha\dot\alpha}=
\begin{array}{||cc||}
1 & 0\\
0 & -1
\end{array}\,.
\end{equation}

The following useful relations are used
\be
\label{rel1}
T_{\alpha\beta\gamma\dots}-T_{\beta\alpha\gamma\dots}=\varepsilon_{\alpha\beta}T^{\delta}
{}_{\delta\gamma\dots},
\ee
\be
\label{rel2}
(\sigma^m\bar{\sigma}^n+\sigma^n\bar{\sigma}^m)_{\alpha}{}^{\beta}=2\eta^{nm}\delta_{\alpha}{}^{\beta}, \quad
(\bar{\sigma}^m\sigma^n+\bar{\sigma}^n\sigma^m)^{\dot\alpha}{}_{\dot\beta}=2\eta^{nm}\delta^{\dot\alpha}{}_{\dot\beta},
\ee
\be
\label{rel3}
(\sigma^m\bar{\sigma}^n)_{\alpha}{}^{\beta}\delta_{\beta}{}^{\alpha}=2\eta^{nm},
\ee
\be
\label{rel4}
(\sigma^m)_{\alpha\dot\alpha}(\bar{\sigma}_m)^{\dot\beta\beta}=2\delta_{\alpha}{}^{\beta}\delta_{\dot\alpha}{}^{\dot\beta},
\ee
\be
\label{rel5}
\sigma^a\bar{\sigma}^b\sigma^c+\sigma^c\bar{\sigma}^b\sigma^a=-2(\eta^{ac}\sigma^b-\eta^{bc}\sigma^a-\eta^{ab}\sigma^c),
\ee
\be
\label{rel6}
\bar{\sigma}^a{\sigma}^b\bar{\sigma}^c+\bar{\sigma}^c{\sigma}^b\bar{\sigma}^a=-2(\eta^{ac}\bar{\sigma}^b-\eta^{bc}\bar{\sigma}^a-\eta^{ab}\bar{\sigma}^c).
\ee

Bosonic coordinates $x^m$  combined with fermionic coordinates $\theta^{\mu},
\bar\theta^{\dot\mu}$  constitute superspace coordinates
 $z^M\sim (x^m,\theta^{\mu},\bar\theta^{\dot\mu})$. Base superspace indices
 are denoted by upper case letters from the middle of
Latin alphabet, while fiber superspace indices are from the beginning of the alphabet.
We use the standard convention for commutation rules
of supercoordinates, exterior product and exterior differentiation in
superspace \cite{wb}, allowing to keep the standard rules of
differentiation and  multiplication of differential forms in superspace.

We also use notations for the following linear spaces over $\mathbb C$:
\begin{itemize}
\item  $\mathbf{U}^k_{p}$
is a space of differential $p$-forms $t^{a(k)}$ such that
$t_b{}^{ba(k-2)}=0$;
\item $\mathbf{V}^k_{p}$ is a space
of differential $p$-forms $\tau_{\alpha}{}^{a(k)}$ such that
$\tau_{\alpha b}{}^{ba(k-2)}=0$, $(\bar\sigma_b)^{\dot\alpha\alpha}\tau_{\alpha}{}^{ba(k-1)}=0$;
\item $\mathbf{\dot V}^k_{p}$ is a space
of differential $p$-forms $\tau^{\dot\alpha a(k)}$ such that
$\tau^{\dot\alpha}{}_b{}^{ba(k-2)}=0$, $(\sigma_b)_{\alpha\dot\alpha}\tau^{\dot\alpha b a(k-1)}=0$;
\item  $\mathbf{U}^{k,1}_{p}$
is a space of differential $p$-forms $t^{a(k),b}$, $k\ge 1$ such that $t^{a(k),a}=0$,
$t_c{}^{ca(k-2),b}=0$ with other traces vanishing as a consequence of the first two conditions;
\item $\mathbf{V}^{k,1}_{p}$ is a space
of differential $p$-forms $\tau_{\alpha}{}^{a(k),b}$, $k\ge 1$ such that $\tau_{\alpha}{}^{a(k),a}=0$,
$\tau_{\alpha c}{}^{ca(k-2),b}=0$, $(\bar\sigma_c)^{\dot\alpha\alpha}\tau_{\alpha}{}^{ca(k-1),b}=0$
with other traces and $\sigma$-longitudinal projections vanishing as a consequence of the first
three conditions.

\end{itemize}

\section*{Appendix B: Technicalities}
\setcounter{equation}{0}\setcounter{section}{2}

The detailed expressions for cell operators introduced in Section \ref{sigmawz} are:
\be
\notag
(\pi_{\downarrow}t)^{a(k-1)}=E_{b}t^{a(k-1)b},
\ee
\be
\notag
(\pi_{\downarrow}\tau)_{\alpha}{}^{a(k-1)}=E_{b}\tau_{\alpha}{}^{a(k-1)b},
\ee
\be
\notag
(\pi_{\uparrow}t)^{a(k+1)}=E^at^{a(k)}-\frac{k}{2(k+1)}E_{b}t^{a(k-1)b}\eta^{aa},
\ee
\be
\notag
(\pi_{\uparrow}\tau)_{\alpha}{}^{a(k+1)}=E^a\tau_{\alpha}{}^{a(k)}-\frac{k}{2(k+1)}E_{b}\tau_{\alpha}{}^{a(k-1)b}\eta^{aa}+\frac{1}{2(k+1)}E^b(\sigma_b)_{\alpha\dot\alpha}(\bar{\sigma}^a)^{\dot\alpha\beta}\tau_{\beta}{}^{a(k)},
\ee
\be
\notag
(\pi^{\times}t)^{a(k)}=E_{b}t^{a(k),b},
\ee
\be
\notag
(\pi^{\times}\tau)_{\alpha}{}^{a(k)}=E_{b}\tau_{\alpha}{}^{a(k),b},
\ee
\be
\notag
(\pi_{\searrow}\tau)^{a(k)}=E_{\dot\alpha}\tau^{\dot\alpha a(k)},
\ee
\be
\notag
(\pi_{\searrow}t)_{\alpha}{}^{a(k-1)}=E^{\dot\alpha}(\sigma_b)_{\alpha\dot\alpha}t^{a(k-1)b},
\ee
\be
\notag
(\pi_{\nearrow}\tau)^{a(k)}=E^{\alpha}(\sigma^a)_{\alpha\dot\alpha}\tau^{\dot\alpha a(k-1)},
\ee
\be
\notag
(\pi_{\nearrow}t)_{\alpha}{}^{a(k)}=E_{\alpha}t^{a(k)}-\frac{k}{2(k+1)}E_{\beta}
(\sigma^a)_{\alpha\dot\alpha}(\bar{\sigma}_b){}^{\dot\alpha\beta}t^{a(k-1)b},
\ee
\be
\notag
(\pi_{\rightarrow}\tau)_{\alpha}{}^{a(k)}=E^b(\sigma_b)_{\alpha\dot\alpha}\tau^{\dot\alpha a(k)}-
\frac{k}{k+1}E_b(\sigma^a)_{\alpha\dot\alpha}\tau^{\dot\alpha a(k-1)b},
\ee
\be
\notag
(\pi_{\swarrow}\tau)^{a(k)}=E^{\alpha}\tau_{\alpha}{}^{a(k)},
\ee
\be
\notag
(\pi_{\nwarrow} \tau)^{ a(k)}=E_{\dot\alpha}(\bar{\sigma}^a)^{\dot\alpha\alpha}\tau_{\alpha}{}^{ a(k-1)},
\ee
\be
\notag
(\pi^{\times}_{\searrow}t)_{\alpha}{}^{a(k-1)}=E^{\gamma}(\sigma_b)_{\gamma\dot\gamma}(\bar{\sigma}_c)^{\dot\gamma\beta}\varepsilon_{\beta\alpha}t^{a(k-1)b,c},
\ee
\be
\notag
(\pi^{\times}_{\nearrow}t)_{\alpha}{}^{a(k)}=E^{\dot\alpha}(\sigma_b)_{\alpha\dot\alpha}t^{a(k),b}-\frac{k}{2(k+1)}
E^{\dot\alpha}(\sigma^a)_{\alpha\dot\beta}
(\bar\sigma_c)^{\dot\beta\beta}(\sigma_b)_{\beta\dot\alpha}t^{a(k-1)c,b}.
\ee
This is the full list of operators such that their image  belongs
to $\mathbf{U}^k_{p}$ and $\mathbf{V}^k_{p}$,
 $p\ge 1$. In  Section \ref{nonzerom} we use operators, that act to
 $\mathbf{\dot V}^k_{p}$ and have the form
\be
\notag
(\pi_{\swarrow}t)^{\dot\alpha a(k-1)}=\phi_{\alpha}(\bar\sigma_m)^{\dot\alpha\alpha}t^{a(k-1)m},
\ee
\be
\notag
(\pi_{\nwarrow}t)^{\dot\alpha a(k)}=\bar\phi^{\dot\alpha}t^{a(k)}-\frac{k}{2(k+1)}
\bar\phi^{\dot\beta}(\bar\sigma^a)^{\dot\alpha\alpha}(\sigma_b)_{\alpha\dot\beta}t^{a(k-1)b},
\ee
\be
\notag
(\pi_{\leftarrow}\tau)^{\dot\alpha a(k)}=e^m(\bar\sigma_m)^{\dot\alpha\alpha}\tau_{\alpha}{}^{a(k)}-
\frac{k}{k+1}e_m(\bar\sigma^a)^{\dot\alpha\alpha}\tau_{\alpha}{}^{a(k-1)m},
\ee
\be
\notag
(\pi_{\downarrow}\tau)^{\dot\alpha a(k-1)}=e_m\tau^{\dot\alpha a(k-1)m},
\ee
\be
\notag
(\pi_{\uparrow}\tau)^{\dot\alpha a(k+1)}=e^a\tau^{\dot\alpha a(k)}-
\frac{k}{2(k+1)}e_m\tau^{\dot\alpha a(k-1)m}\eta^{aa}+
\frac{1}{2(k+1)}e^m(\bar\sigma_m)^{\dot\alpha\alpha}(\sigma^a)_{\alpha\dot\beta}\tau^{\dot\beta a(k)}.
\ee

The compatibility conditions for (\ref{m1})-(\ref{m4}) imply
\be
\notag
\{D,\sigma_{\downarrow}\}C+\{\sigma_{\swarrow},\sigma_{\searrow}\}C=0 \quad \Rightarrow
\quad -2iU_{\downarrow}(k)+V_{\swarrow}(k-1)U_{\searrow}(k)+W_{\searrow}(k-1)U_{\swarrow}(k)=0,
\ee
\be
\notag
\{D,\sigma_{\uparrow}\}C+\{\sigma_{\nwarrow},\sigma_{\nearrow}\}C=0 \quad \Rightarrow
\quad -2iU_{\uparrow}(k)+V_{\nwarrow}(k)U_{\nearrow}(k)+W_{\nearrow}(k)U_{\nwarrow}(k)=0,
\ee
\be
\notag
\{\sigma_{\downarrow},\sigma_{\uparrow}\}C+\{\sigma_{\nearrow},\sigma_{\swarrow}\}C+
\{\sigma_{\nwarrow},\sigma_{\searrow}\}C=0 \quad \Rightarrow
\quad U_{\downarrow}(k+1)U_{\uparrow}(k)\frac{k}{k+1}\frac{k+2}{k+1}=U_{\uparrow}(k-1)
U_{\downarrow}(k),
\ee
\be
\notag
W_{\nearrow}(k-1)U_{\swarrow}(k)=\frac{k}{2(k+1)}V_{\swarrow}(k)U_{\nearrow}(k),
\quad V_{\nwarrow}(k-1)U_{\searrow}(k)=\frac{k}{2(k+1)}W_{\searrow}(k)U_{\nwarrow}(k),
\ee
\be
\notag
\{\sigma_{\downarrow},\sigma_{\searrow}\}C=0 \quad \Rightarrow
\quad V_{\downarrow}(k-1)U_{\searrow}(k)=U_{\searrow}(k-1)U_{\downarrow}(k),
\ee
\be
\notag
\{\sigma_{\uparrow},\sigma_{\nearrow}\}C=0 \quad \Rightarrow
\quad V_{\uparrow}(k)U_{\nearrow}(k)\frac{k+2}{k+1}=U_{\nearrow}(k+1)U_{\uparrow}(k),
\ee
\be
\notag
(\{\sigma_{\downarrow},\sigma_{\nearrow}\}+\sigma_{\rightarrow}\sigma_{\swarrow})C=0
\quad \Rightarrow
\quad V_{\downarrow}(k)U_{\nearrow}(k)=U_{\nearrow}(k-1)U_{\downarrow}(k),
\ee
\be
\notag
V_{\downarrow}(k)U_{\nearrow}(k)\frac{1}{2(k+1)}=W_{\rightarrow}(k-1)U_{\swarrow}(k),
\ee
\be
\notag
(\{\sigma_{\uparrow},\sigma_{\searrow}\}+\sigma_{\rightarrow}\sigma_{\nwarrow})C=0\quad
\Rightarrow
\quad U_{\searrow}(k+1)U_{\uparrow}(k)=(k+1)W_{\rightarrow}(k)U_{\nwarrow}(k),
\ee
\be
\notag
V_{\uparrow}(k-1)U_{\searrow}(k)=W_{\rightarrow}(k)U_{\nwarrow}(k)\frac{k^2(k+2)}{(k+1)^2},
\ee
\be
\notag
\{\sigma_{\downarrow},\sigma_{\swarrow}\}C=0 \quad \Rightarrow
\quad V_{\downarrow}(k-1)U_{\swarrow}(k)=U_{\swarrow}(k-1)U_{\downarrow}(k),
\ee
\be
\notag
\{\sigma_{\uparrow},\sigma_{\nwarrow}\}C=0 \quad \Rightarrow
\quad V_{\uparrow}(k)U_{\nwarrow}(k)\frac{k+2}{k+1}=U_{\nwarrow}(k+1)U_{\uparrow}(k),
\ee
\be
\notag
(\{\sigma_{\downarrow},\sigma_{\nwarrow}\}+\sigma_{\leftarrow}\sigma_{\searrow})C=0
\quad \Rightarrow
\quad V_{\downarrow}(k)U_{\nwarrow}(k)=U_{\nwarrow}(k-1)U_{\downarrow}(k),
\ee
\be
\notag
V_{\downarrow}(k)U_{\nwarrow}(k)\frac{1}{2(k+1)}=V_{\leftarrow}(k-1)U_{\searrow}(k),
\ee
\be
\notag
\{\sigma_{\uparrow},\sigma_{\swarrow}\}+\sigma_{\leftarrow}(\sigma_{\nearrow})C=0\quad
\Rightarrow
\quad U_{\swarrow}(k+1)U_{\uparrow}(k)=(k+1)V_{\leftarrow}(k)U_{\nearrow}(k),
\ee
\be
\notag
V_{\uparrow}(k-1)U_{\swarrow}(k)=V_{\leftarrow}(k)U_{\nearrow}(k)\frac{k^2(k+2)}{(k+1)^2},
\ee
\be
\notag
\{\sigma_{\downarrow},\sigma_{\swarrow}\}\chi=0 \quad \Rightarrow
\quad U_{\downarrow}(k)V_{\swarrow}(k)=V_{\swarrow}(k-1)V_{\downarrow}(k),
\ee
\be
\notag
\{\sigma_{\downarrow},\sigma'_{\swarrow}\}\chi=0 \quad \Rightarrow
\quad T_{\downarrow}(k)V'_{\swarrow}(k)=V'_{\swarrow}(k-1)V_{\downarrow}(k),
\ee
\be
\notag
 \{\sigma_{\uparrow},\sigma_{\nwarrow}\}\chi=0 \quad \Rightarrow
\quad U_{\uparrow}(k+1)V_{\nwarrow}(k)=\frac{k+2}{k+1}V_{\nwarrow}(k+1)V_{\uparrow}(k),
\ee
\be
\notag
 \{\sigma_{\uparrow},\sigma'_{\nwarrow}\}\chi=0 \quad \Rightarrow
\quad T_{\uparrow}(k+1)V'_{\nwarrow}(k)=\frac{k+2}{k+1}V'_{\nwarrow}(k+1)V_{\uparrow}(k),
\ee
\be
\notag
(\{\sigma_{\downarrow},\sigma_{\nwarrow}\}+\sigma_{\searrow}\sigma_{\leftarrow})\chi=0 \quad
\Rightarrow
\quad U_{\downarrow}(k+1)V_{\nwarrow}(k)\frac 1{k+1}=W_{\searrow}(k)V_{\leftarrow}(k),
\ee
\be
\notag
U_{\downarrow}(k+1)V_{\nwarrow}(k)\frac{k(k+2)}{(k+1)^2}=V_{\nwarrow}(k-1)V_{\downarrow}(k),
\ee
\be
\notag
(\{\sigma_{\downarrow},\sigma'_{\nwarrow}\}+\sigma'_{\searrow}\sigma_{\leftarrow})\chi=0 \quad
\Rightarrow
\quad T_{\downarrow}(k+1)V'_{\nwarrow}(k)\frac 1{k+1}=W'_{\searrow}(k)V_{\leftarrow}(k),
\ee
\be
\notag
T_{\downarrow}(k+1)V'_{\nwarrow}(k)\frac{k(k+2)}{(k+1)^2}=V'_{\nwarrow}(k-1)V_{\downarrow}(k),
\ee
\be
\notag
(\{\sigma_{\uparrow},\sigma_{\swarrow}\}+\sigma_{\nearrow}\sigma_{\leftarrow})\chi=0 \quad
\Rightarrow
\quad V_{\swarrow}(k+1)V_{\uparrow}(k)\frac{1}{2(k+1)}=W_{\nearrow}(k)V_{\leftarrow}(k),
\ee
\be
\notag
U_{\uparrow}(k)V_{\swarrow}(k)=\frac{k+2}{k+1}V_{\swarrow}(k+1)V_{\uparrow}(k),
\ee
\be
\notag
(\{\sigma_{\uparrow},\sigma'_{\swarrow}\}+\sigma'_{\nearrow}\sigma_{\leftarrow})\chi=0 \quad
\Rightarrow
\quad V'_{\swarrow}(k+1)V_{\uparrow}(k)\frac{1}{2(k+1)}=W'_{\nearrow}(k)V_{\leftarrow}(k),
\ee
\be
\notag
T_{\uparrow}(k)V'_{\swarrow}(k)=\frac{k+2}{k+1}V'_{\swarrow}(k+1)V_{\uparrow}(k),
\ee
\be
\notag
(\{D,\sigma_{\downarrow}\}+\sigma_{\searrow}\sigma_{\swarrow}+\sigma'_{\searrow}\sigma'_{\swarrow})\chi=0 \quad \Rightarrow
\quad -2iV_{\downarrow}(k)+U_{\searrow}(k)V_{\swarrow}(k)+
T_{\searrow}(k)V'_{\swarrow}(k)=0,
\ee
\be
\notag
(\{D,\sigma_{\uparrow}\}+\sigma_{\nearrow}\sigma_{\nwarrow}+\sigma'_{\nearrow}\sigma'_{\nwarrow})\chi=0 \quad \Rightarrow
\quad -2iV_{\uparrow}(k)+\frac{k+1}{k+2}(U_{\nearrow}(k+1)V_{\nwarrow}(k)+T_{\nearrow}(k+1)V'_{\nwarrow}(k))=0,
\ee
\be
\notag
(\{\sigma_{\downarrow},\sigma_{\uparrow}\}+\sigma_{\rightarrow}\sigma_{\leftarrow}+
\sigma_{\searrow}\sigma_{\nwarrow}+\sigma_{\nearrow}\sigma_{\swarrow}
+\sigma'_{\searrow}\sigma'_{\nwarrow}+\sigma'_{\nearrow}\sigma'_{\swarrow})\chi=0 \Rightarrow
\ee
\be
\notag
U_{\searrow}(k+1)V_{\nwarrow}(k)+T_{\searrow}(k+1)V'_{\nwarrow}(k)=0,
\quad  U_{\nearrow}(k)V_{\swarrow}(k)+T_{\nearrow}(k)V'_{\swarrow}(k)=0,
\ee
\be
\notag
W_{\rightarrow}(k)V_{\leftarrow}(k)=\frac{1}{2(k+1)^2}V_{\downarrow}(k+1)V_{\uparrow}(k),
\quad  V_{\downarrow}(k+1)V_{\uparrow}(k)=\frac{(k+1)^4}{k^2(k+2)^2}V_{\uparrow}(k-1)V_{\downarrow}(k),
\ee
\be
\notag
(\{D,\sigma_{\leftarrow}\}+\{\sigma_{\swarrow},\sigma_{\nwarrow}\}+\{\sigma'_{\swarrow},\sigma'_{\nwarrow}\})\chi=0 \quad \Rightarrow
\ee
\be
\notag
-4iV_{\leftarrow}(k)-2U_{\swarrow}(k+1)V_{\nwarrow}(k)+U_{\nwarrow}(k)V_{\swarrow}(k)
-2T_{\swarrow}(k+1)V'_{\nwarrow}(k)+T_{\nwarrow}(k)V'_{\swarrow}(k)=0,
\ee
\be
\notag
(\{\sigma_{\downarrow},\sigma_{\leftarrow}\}+\sigma_{\swarrow}\sigma_{\swarrow}+\sigma'_{\swarrow}\sigma'_{\swarrow})\chi=0 \quad
\Rightarrow
\quad V_{\leftarrow}(k-1)V_{\downarrow}(k)=W_{\downarrow}(k)V_{\leftarrow}(k)\frac{k+2}{k+1},
\ee
\be
\notag
U_{\swarrow}(k)V_{\swarrow}(k)+T_{\swarrow}(k)V'_{\swarrow}(k)=0,
\ee
\be
\notag
(\{\sigma_{\uparrow},\sigma_{\leftarrow}\}+\sigma_{\nwarrow}\sigma_{\nwarrow}+\sigma'_{\nwarrow}\sigma'_{\nwarrow})\chi=0 \quad \Rightarrow
\quad W_{\uparrow}(k)V_{\leftarrow}(k)=V_{\leftarrow}(k+1)V_{\uparrow}(k)\frac{k+3}{k+2},
\ee
\be
\notag
U_{\nwarrow}(k+1)V_{\nwarrow}(k)+T_{\nwarrow}(k+1)V'_{\nwarrow}(k)=0.
\ee

\end{document}